\documentclass[12pt, a4paper]{article}
\usepackage{amsmath,amssymb}
\usepackage{graphicx}

\newtheorem{theorem}{Theorem}

\begin{document}

\title{On linear degeneracy  of integrable quasilinear systems in higher dimensions}
\author{E.V. Ferapontov, K.R. Khusnutdinova and C. Klein$^*$
  }
   \date{}
   \maketitle
   \vspace{-7mm}
\begin{center}
Department of Mathematical Sciences \\ 
Loughborough University \\
Loughborough, Leicestershire LE11 3TU \\ United Kingdom \\[2ex]
$^*$ Institut de Math\'ematiques de Bourgogne\\
9 avenue Alain Savary, BP 47870\\
21078 Dijon Cedex, France \\[2ex]
e-mails: \\[1ex] \texttt{E.V.Ferapontov@lboro.ac.uk}\\
\texttt{K.Khusnutdinova@lboro.ac.uk}\\
\texttt{Christian.Klein@u-bourgogne.fr}

\end{center}

\bigskip

\begin{abstract}

We investigate $(d+1)$-dimensional quasilinear systems which are integrable by the method of hydrodynamic reductions. In the case $d\geq 3$ we formulate a conjecture that any such system with an irreducible dispersion relation must be linearly degenerate. We prove this conjecture in the $2$-component case, providing a complete classification of multi-dimensional integrable systems in question. In particular, our results imply the non-existence of $2$-component integrable systems of hydrodynamic type for $d\geq 6$. 

In the second half of the paper we  discuss a numerical and analytical evidence for the impossibility of the breakdown of smooth initial data for    linearly degenerate  systems in $2+1$ dimensions.  

\bigskip
MSC: 35L40, 35L65, 37K10.

\bigskip
Keywords: Multi-dimensional Quasilinear Systems, Hydrodynamic Reductions, Integrability, Linear Degeneracy,
Cauchy Problem, Classical Solutions.
\end{abstract}

\newpage

\section{Introduction}

In this paper we discuss  $(d+1)$-dimensional quasilinear systems of the form
\begin{equation}
{\bf u}_t+\sum_{i=1}^d A_{i}({\bf u}) {\bf u}_{x^i}=0,
\label{1}
\end{equation}
where $t, x^i$ are independent variables,  ${\bf u}$ is  an
$n$-component column vector and $A_i({\bf u})$ are $n\times
n$ matrices. Systems of this type play an important role in continuum mechanics and physics, and there is a vast literature covering their analytical, geometric and applied aspects, see e.g.  \cite{Dafermos, Majda, R2, Serre, Godunov, Dub, Tsarev}. In this paper we  assume that the dispersion relation  $\det(\mu I+\alpha^i A_i)=0$ defines an irreducible algebraic variety (for $n=2$, an irreducible quadric) in the variables $\mu, \ \alpha^i$. Our prime goal is the classification of {\it integrable} systems of the form (\ref{1}) in any dimension. 

For $d=1$, any  two-component system of the form
$$
{\bf u}_t+A({\bf u}) {\bf u}_{x}=0
$$
is linearizable by a hodograph transformation which interchanges the dependent and independent variables and, therefore, is automatically integrable. In particular, any such system possesses an infinity of conservation laws and commuting flows. 

For $d=2$,
\begin{equation}
{\bf u}_t+A({\bf u}) {\bf u}_{x}+B({\bf u}) {\bf u}_{y}=0,
\label{2+1}
\end{equation}
 the situation is more subtle. Even the very concept of   integrability for multi-dimensional systems of the form (\ref{1}) requires an alternative approach. Such approach, based on the method of {\it hydrodynamic reductions}, was outlined in \cite{Fer3}. It was proposed to call a $(2+1)$-dimensional system integrable if it possesses infinitely many multi-phase solutions of the form ${\bf u}={\bf u}(R^1, ..., R^N)$ where the phases 
 $R^i(x, y, t)$ satisfy a pair of commuting hydrodynamic type systems 
 \begin{equation}
 R^i_t=\lambda^i(R)\ R^i_x, ~~~~ R^i_y=\mu^i(R)\ R^i_x;
\label{Rim}
\end{equation}
we emphasize that the number of phases $N$ is allowed to be arbitrary. In what follows, we will refer to the equations (\ref{Rim}) as $N$-component hydrodynamic reductions. Multi-phase solutions of this type, known in gas dynamics as nonlinear interactions of planar simple waves, date back to \cite{Sidorov,  Burnat3, Perad1,   Grundland}. Their first appearance in the context of integrable systems and, in particular, the dispersionless Kadomtsev-Petviashvili hierarchy, is due to \cite{Gibb94, GibTsa96}. The requirement of the existence of multi-phase solutions imposes strong constraints on the matrices $A({\bf u})$ and $B({\bf u})$. In the two-component case these integrability conditions were derived in \cite{Fer4}, and are included in Sect. 4.1 of this paper. In particular, they imply that any integrable system of this kind is necessarily conservative, and possesses a dispersionless Lax pair. It was demonstrated in \cite{Odesskii1, Odesskii2} that the integrability conditions  can be resolved in generalized hypergeometric functions. 
 
Remarkably,  for $d\geq3$ the situation simplifies dramatically:  the requirement of the integrability proves to be too strong, so that very few examples survive in many dimensions. According to the definition proposed in \cite{Fer5}, a $(d+1)$-dimensional system of the form (\ref{1}) is said to be integrable if it possesses infinitely many $N$-component hydrodynamic reductions parametrized by $(d-1)N$ arbitrary functions of one variable. Here by a hydrodynamic reduction we mean a decomposition of a $(d+1)$-dimensional system into $d$ commuting systems of the form (\ref{Rim}).  Although the derivation of the integrability conditions based on the method of hydrodynamic reductions leads to a quite complicated analysis, there exists a simple way to bypass lengthy calculations. It is based on the requirement that {\it all} traveling wave reductions of a $(d+1)$-dimensional integrable system to  $(2+1)$-dimensional systems should be themselves integrable. Since the integrability conditions in $2+1$ dimensions are explicitly known, this provides a set of strong constraints which are therefore necessary for the integrability. In fact, in our case they turn out  to be sufficient. Our main result is the following

\begin{theorem} 
There exist no two-component integrable systems of the form (\ref{1}) in dimensions $6+1$ and higher.  In $5+1$ dimensions, 
there exists a unique integrable system. Up to linear transformations of the independent variables, it can be brought to the form
\begin{equation}
v_{t}=w_{ x}+wv_{ r}-vw_{ r},  ~~~~ w_{ z}=v_{ y}+vw_{ s}-wv_{ s}.
\label{ma}
\end{equation}
In $4+1$  and $3+1$  dimensions, any two-component integrable system   can be obtained as a traveling wave reduction of the system (\ref{ma}). 

\end{theorem}
The proof of this result is given in Sect. 4.2. The system (\ref{ma}) first appeared in \cite{Fer5}, see Sect. 2 for more details. An important property of this example is its {\it linear degeneracy}. Recall that a matrix $A({\bf u})$ is said to be linearly degenerate if its eigenvalues (assumed real and distinct) are constant in the direction of the corresponding  eigenvectors. Explicitly, $L_{r^i}\lambda^i=0$, no summation, where $L_{r^i}$ is the Lie derivative in the direction of the  eigenvector $r^i$, and $A r^i=\lambda^i r^i$. A system of the form (\ref{1}) is said to be linearly degenerate (totally linearly degenerate in the terminology of \cite{Majda}), if any matrix 
$A({\bf u})=\alpha^i A_i({\bf u})$ is linearly degenerate for any values of the constants $\alpha^i$. Theorem 1 and other existing examples support the following conjecture.

\medskip

\noindent  {\bf Conjecture.} {\it For $d\geq 3$, any $n$-component $(d+1)$-dimensional integrable system of the form (\ref{1}) with an irreducible dispersion relation  must be linearly degenerate.}

\medskip

According to the conjecture of Majda \cite{Majda}, p. 89,  linearly degenerate systems are quite exceptional from the point of view of the global solvability of the Cauchy problem: the shock wave formation, which is a standard scenario for genuinely nonlinear systems, never happens for smooth initial data. Although this fact is well-established in $1+1$ dimensions  \cite{R1, R2, Liu, Serre}, the multi-dimensional situation is still out of  reach. We refer to the recent work \cite{Manakov1, Manakov2, Manakov3} which utilizes a novel version of the inverse scattering transform to investigate the breakdown phenomena for some $(2+1)$-dimensional dispersionless integrable models. In Sect. 3 we provide some further analytical and numerical evidence in support of Majda's conjecture. As an example, we choose a particular $(2+1)$-dimensional traveling wave reduction of the system (\ref{ma}), namely, 
\begin{equation}
v_t+w_x=0, ~~~~ w_t+w_y+vw_x-wv_x=0,
\label{max}
\end{equation}
see Sect. 3.1 for the discussion of this example. In a somewhat different form, it appeared in \cite{Shabat1}. In Sections 3.2--3.3, applying  the method of hydrodynamic reductions and other simple techniques,  we construct some global exact solutions of the system 
 (\ref{max}). In Sect. 3.4 we summarize the results of numerical simulations based on the  Friedrichs-symmetrized representation of the system (\ref{max}). These results clearly indicate that hump-like initial data tend to spread out, and do not break down in finite time.

\section{Examples of  multi-dimensional integrable  systems}

In this section we list some examples of integrable $(d+1)$-dimensional quasilinear systems of the form (\ref{1}) for
$d\geq 3$. All  such systems  turn out to be linearly degenerate, namely, 
 an arbitrary matrix of the linear pencil $A({\bf u})=\alpha^iA_i({\bf u})$, where $\alpha^i$ are arbitrary constants,  must be linearly degenerate (totally linearly degenerate systems in the terminology of \cite{Majda}). We recall that there exists a useful criterion of linear degeneracy  which can be summarized as follows. Given a matrix $A({\bf u})$, let us introduce its characteristic polynomial,
 $$
 det(\lambda E-A({\bf u}))={\lambda  }^n  +  f_1({\bf u}){\lambda
}^{n-1} +f_2({\bf u}){\lambda}^{n-2}+ \ldots + f_n({\bf u}).
$$
The condition of linear degeneracy is  given by \cite{Fer},
\begin{equation}
\nabla f_1~A^{n-1}+\nabla  f_2~A^{n-2}+\ldots  +\nabla  f_n=0,
\label{ldeg}
\end{equation}
where $\nabla$ is the operator of the gradient,  $\nabla f=({{\partial
f}\over {\partial u^1}},\ldots , {{\partial f}\over {\partial
u^n}})$, and $A^k$ denotes  $k$-th power of the matrix $A$.
Our first example plays a key role in the classification of two-component multi-dimensional quasilinear systems.

\medskip

\noindent {\bf Example 1.} The six-dimensional two-component integrable system (\ref{ma}), 
$$
v_{t}=w_{ x}+wv_{ r}-vw_{ r},  ~~~~ w_{ z}=v_{ y}+vw_{ s}-wv_{ s},
$$
first appeared in \cite{Fer5}. It can also be obtained  as  the condition of commutativity  of two parameter-dependent vector fields,
$$
[\partial_ { z}-v\partial_{ s}-\lambda \partial_{ x}+\lambda v \partial_{ r}, ~~
\partial_{ y}-w\partial_{ s}-\lambda \partial_{ t}+\lambda w \partial_{ r}]=0.
$$
It was demonstrated in \cite{Fer5} that  system ({\ref{ma}) possesses infinitely many $n$-component hydrodynamic reductions parametrized by $4n$ arbitrary functions of one variable. This example is a result of the following construction.
In  \cite{Fer4},  we proposed a  characterization of  two-component $(2+1)$-dimensional integrable systems of the form
$$
\left(\begin{array}{c}
v\\
w
\end{array}\right)_t+A(v, w)\left(\begin{array}{c}
v\\
w
\end{array}\right)_x+B(v, w)\left(\begin{array}{c}
v\\
w
\end{array}\right)_y=0.
$$
The integrability conditions constitute a complicated overdetermined system of second order PDEs for the $2\times 2$ matrices $A$ and $B$ which are presented in Sect. 4.1. 
In the particular case when  $A$ is taken to be 
linearly degenerate,
\begin{equation}
 A=\left(
\begin{array}{cc}
w & 0 \\
\ \\
0 & v
\end{array}
\right),
 \label{lindeg}
 \end{equation}
 the integrability conditions imply the following generic expression for $B$,
\begin{equation}
 B=\left(
\begin{array}{cc} 
 \frac{f(w)}{w-v}-\alpha w^2 &  \frac{f(v)}{w-v} \\ 
 \ \\
\frac{f(w)}{v-w} & \frac{f(v)}{v-w}-\alpha v^2
 \end{array}
 \right),
 \label{lindeg1}
 \end{equation}
where $f$ is a cubic polynomial, $f(v)=\alpha v^3+\beta v^2 +\gamma v 
+\delta$, and $\alpha, \beta, \gamma, \delta$ are arbitrary 
constants. A remarkable property of this example is that
{\it any} matrix in the linear pencil $B+\mu A$ is also linearly 
degenerate. Explicitly, one has
 $$
 \begin{array}{c}
  B=\delta B_1+\gamma B_2+\beta B_3+\alpha B_4= \\
 \ \\
 \delta \left(
\begin{array}{cc} 
 \frac{1}{w-v} &  \frac{1}{w-v} \\ 
\frac{1}{v-w} & \frac{1}{v-w}
 \end{array} \right)+
 \gamma \left(
\begin{array}{cc} 
 \frac{w}{w-v} &  \frac{v}{w-v} \\ 
\frac{w}{v-w} & \frac{v}{v-w}
 \end{array}
 \right)+
 \beta \left(
\begin{array}{cc} 
 \frac{w^2}{w-v} &  \frac{v^2}{w-v} \\ 
\frac{w^2}{v-w} & \frac{v^2}{v-w}
 \end{array}
 \right)+
 \alpha \left(
\begin{array}{cc} 
 \frac{vw^2}{w-v} &  \frac{v^3}{w-v} \\ 
\frac{w^3}{v-w} & \frac{wv^2}{v-w}
 \end{array}
 \right).
 \end{array}
$$
Let us introduce the $(5+1)$-dimensional system
$$
\left(\begin{array}{c}
v\\
w
\end{array}\right)_t+A\left(\begin{array}{c}
v\\
w
\end{array}\right)_x+ \\
\ \\
B_1\left(\begin{array}{c}
v\\
w
\end{array}\right)_y+
B_2\left(\begin{array}{c}
v\\
w
\end{array}\right)_z+
B_3\left(\begin{array}{c}
v\\
w
\end{array}\right)_s+
B_4\left(\begin{array}{c}
v\\
w
\end{array}\right)_r
=0.
$$
Notice that an arbitrary linear combination of matrices $A$ and $B_1, B_2, B_3, B_4$ is automatically linearly degenerate. This can be verified directly using the criterion (\ref{ldeg}). In the new variables $V=v+w,\  W=vw$, our system reduces to
$$
V_t+W_x+WV_r-VW_r=0, ~~~
W_t+VW_x-WV_x+V_y+W_z+VW_s-WV_s=0,
$$
taking a fully symmetric form (\ref{ma}) after an obvious linear transformation of the independent variables. We would like to emphasize that there was no a priori reason for the $(5+1)$-dimensional system obtained in this way to be integrable: this is a miracle which is due to the linear degeneracy. 
 Notice that a particular dimensional reduction of the system (\ref{ma}), 
\begin{equation}
m_{t}=n_{ x},  ~~~~ n_{ t}=m_{ y}+mn_{ x}-nm_{ x},
\label{paveq}
\end{equation}
 known as the Pavlov equation, has been extensively discussed in the context of hydrodynamic chains and the `universal hierarchy'  \cite{Pavlov, Shabat1, Shabat2, Manakov3}. 

\medskip

\noindent {\bf Example 2}.  The following four-dimensional three-component integrable system appeared in \cite{d3}  in the context of hyper-Hermitian conformal structures,
$$
A{\bf N}_x+B_1{\bf N}_{y^1}+B_2{\bf N}_{y^2}+B_3{\bf N}_{y^3}=0,
$$
here ${\bf N}$ is a three-component column vector, ${\bf N}=(N_1,  N_2,  N_3)^t$, and $A, B_1, B_2, B_3$ are $3\times 3$ matrices,
 $$
A=\left(\smallmatrix
 1 &  -N_3&N_2 \\ 
N_3 &1&-N_1\\
-N_2&N_1&1
\endsmallmatrix \right),
~~~
B_1= \left(\smallmatrix
0 &  0&0 \\ 
0&0&1\\
0&-1&0
\endsmallmatrix
 \right), ~~ B_2= \left(
\smallmatrix
0&0&-1\\
0&0&0\\
1&0&0
\endsmallmatrix
 \right), ~~ B_3= \left(
\smallmatrix
0&1&0\\
-1&0&0\\
0&0&0
\endsmallmatrix
 \right).
$$
 Multiplying  by $A^{-1}$ one can bring this system to an evolutionary form, moreover,  any linear combination of matrices $A^{-1}B_i$ is linearly degenerate. Notice, however, that the dispersion relation of this system is not irreducible.

A general scheme for constructing  multi-dimensional integrable systems based on the commutativity of vector fields was discussed in \cite{Bogdanov}. A broad class of new  examples of integrable  linearly degenerate systems in $2+1$ dimensions was constructed in \cite{Odesskii3}.

\section{Global solutions of  linearly degenerate systems}

In $1+1$ dimensions, linearly degenerate systems are known to be quite exceptional from the point of view of the solvability of the Cauchy problem: generic smooth initial data do not develop shocks in finite time \cite{R1, R2, Liu, Serre}. The conjecture of Majda
\cite{Majda}, p. 89, suggests that the same statement should be true in higher dimensions, namely, for linearly degenerate systems the shock formation never happens for smooth initial data. For a particular case of the Pavlov equation, this conjecture was confirmed in \cite{Manakov3} based on the novel version of the inverse scattering transform. In this section we discuss another dimensional reduction of the system (\ref{ma}), namely, the system (\ref{max}), 
$$
v_t+w_x=0, ~~~~ w_t+w_y+vw_x-wv_x=0.
$$
In contrast to the Pavlov equation (\ref{paveq}), this system has a well-posed initial value problem. We also  refer to \cite{B1, B2} for a detailed discussion of the $10\times 10$ linearly degenerate Born-Infeld system which is a nonlinear version of Maxwell's equations. We hope that our linearly degenerate example will be more appropriate for establishing the global existence type results.

Some useful properties of system (\ref{max}) are discussed in Sect. 3.1: these include the conservative representation and the Friedrichs-symmetrized form, which is most suitable for numerical simulations.  In Sect. 3.2 and 3.3 we construct  globally defined exact solutions of system (\ref{max}) by reducing it to known integrable equations in $1+1$ dimensions. This includes   reductions to the sinh-Gordon and KdV  equations (Sect. 3.2), as well as the reduction to a pair of $N$-component linearly degenerate systems (Sect. 3.3). Sect. 3.4 contains numerical simulations of the system (\ref{max}) which support the conjecture that smooth initial data do not break down in finite time. 
We emphasize that all these properties are not specific to the system (\ref{max}), and hold for any linearly degenerate integrable system.

\subsection{Properties of System (\ref{max})}

One can show that system ({\ref{max}) possesses exactly three zero order  conservation laws,
$$
\begin{array}{c}
v_t+w_x=0, \\
\ \\
(v^2-w)_t+(vw)_x-w_y=0, \\
\ \\
(1/w)_t+(v/w)_x+(1/w)_y=0.
\end{array}
$$
It can be written in a Friedrichs-symmetrized form,
\begin{equation}
\left(
\begin{array}{cc}
-w & 0 \\
\ \\
0 & 1
\end{array}
\right)
\left(
\begin{array}{c}
v \\
\ \\
w
\end{array}
\right)_t+
\left(
\begin{array}{cc}
0 & -w \\
\ \\
-w & v
\end{array}
\right)
\left(
\begin{array}{c}
v \\
\ \\
w
\end{array}
\right)_x+
\left(
\begin{array}{cc}
0&0 \\
\ \\
0&1
\end{array}
\right)
\left(
\begin{array}{c}
v \\
\ \\
w
\end{array}
\right)_y=0,
\label{max11}
\end{equation}
which shows that one should expect at least finite time existence results in the region $w< 0$. 
 Introducing the variables $p=-2v/w^2, ~ q=1/w^2$, one can rewrite  (\ref{max}) in the Godunov form \cite{Godunov}, 
$$
(f_p)_t+(g_p)_x+(h_p)_y=0, ~~ (f_q)_t+(g_q)_x+(h_q)_y=0, 
$$
where 
$$
f=-\frac{p^2}{4q}-2\sqrt q, ~~~ g=\frac{p}{\sqrt q}, ~~~ h=-2\sqrt q.
$$
\noindent {\bf Remark. }  The system (\ref{max}) possesses a Lax pair which can be represented as the condition of commutativity of two vector fields,
\begin{equation}
[\partial_t+\partial_y+(v+\lambda)\partial_x, \ w\partial_x-\lambda\partial_t]=0.
\label{Lax}
\end{equation}
Lax pairs of this kind form a basis of the novel generalization of the inverse scattering transform \cite{Manakov1, Manakov2, Manakov3}.

\subsection{Reductions to integrable  equations}

The system (\ref{max}) possesses a number of remarkable reductions to soliton equations in $1+1$ dimensions.  Below we describe two particular reductions, to the sinh-Gordon and the KdV equations, respectively. For convenience, we make a change of variables  $x, y, t \to x, y, \tau=t-y$ which brings  system (\ref{max}) into the form
\begin{equation}
v_{\tau}+w_x=0, ~~~~ w_y+vw_x-wv_x=0.
\label{max1}
\end{equation}

\medskip

\noindent {\bf Reduction to the sinh-Gordon equation.}
  Applying the procedure of \cite{d2} to the Lax pair (\ref{Lax}) one can obtain a class of exact solutions of the system (\ref{max1}) in the form
 \begin{equation}
 v=u_y \cos x, ~~~
 w=-\sinh u \sin x+\cosh u,
 \label{sinh}
  \end{equation}
 where $u(y, \tau)$ satisfies the $\sinh$-Gordon  equation
 $$
 u_{y\tau}=\sinh u.
 $$
In the original variables, it takes the form  $ u_{yt}+u_{tt}=\sinh u.$ Taking a traveling wave solution   in the form
 \begin{eqnarray}
&&u = {\rm arccosh} \left ( e - (e-1)\  {\rm sn}^2 [\sigma (y - c t)\   | \ m] \right ) {\rm sign} \left (\cos \frac{\pi (y - ct)}{T}\right ), \nonumber \\
&&T = 2 \sigma^{-1} K(m), \quad \sigma = \sqrt {\frac{e+1}{2 c (1-c)}}, \quad m = \frac{e-1}{e+1}, \label{u}
 \end{eqnarray}
 where ${\rm sn}[ \cdot | m]$ is a Jacobi elliptic function and $K(m)$ is the complete elliptic integral of the first kind (e.g. \cite{Abramowitz}), $0 < c < 1$ and $e>1$, one obtains bounded doubly periodic solutions  $v(x, y, t)=(u_y+u_t) \cos x$ and $
 w(x, y, t)=-\sinh u \sin x+\cosh u$ of the system (\ref{max})
 shown in  Fig.~\ref{fig1}  for $c = 1/3$ and $e=1.1$ (plotted using {\it Mathematica} \cite{Mathematica}). These solutions remain non-singular for all values of $t$. 

  \begin{figure}
[!htbp]
\begin{center}
 \includegraphics[width=65mm]{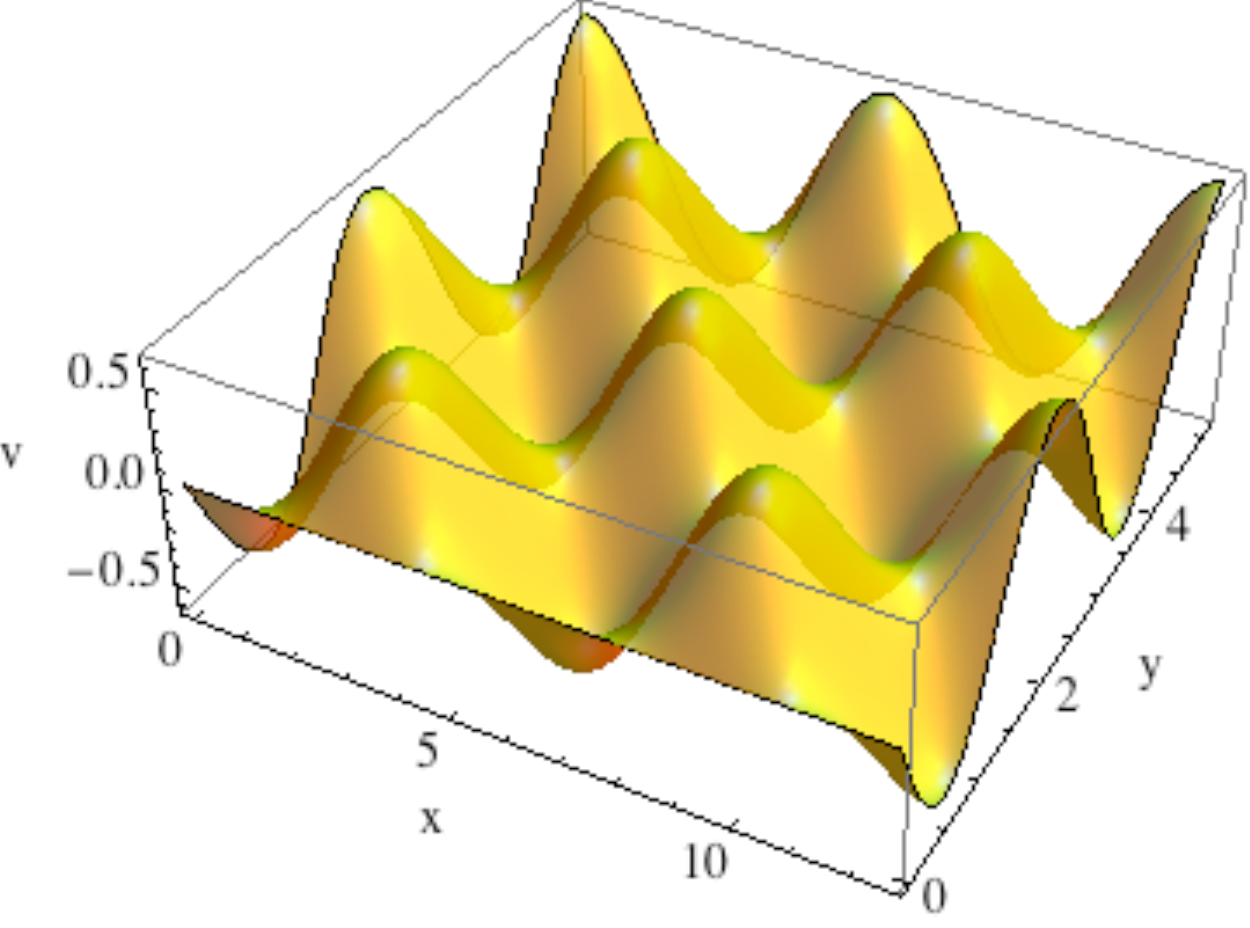} 
 \includegraphics[width=65mm]{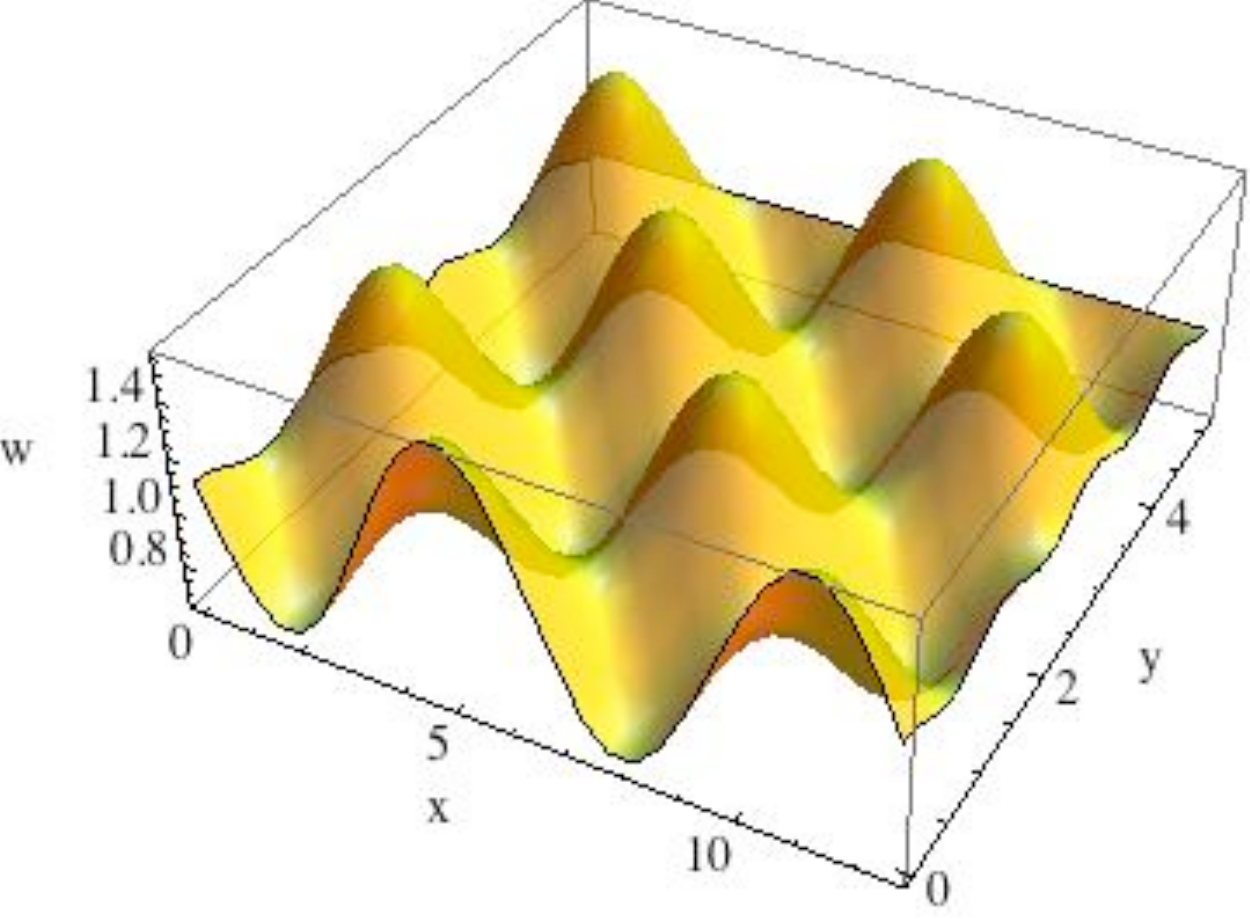} 
\caption{Solution (\ref{sinh}), (\ref{u}) to the system (\ref{max}) for $v$ and $w$  at $t =0$.}
\label{fig1}
\end{center}
\vspace*{-5mm}
\end{figure}

 \medskip
 
 \noindent {\bf Reduction to the KdV equation.}  Let $v(x, y, \tau)$ be a solution to a pair of commuting flows
 $$
 v_{\tau}=(\varphi^2)_x, ~~~ v_y=\frac{1}{4}v_{xxx}-3vv_x,
 $$
 where $\varphi$ satisfies the system
 $$
 \varphi_{xx}=2v\varphi, ~~~ \varphi_y=\frac{1}{2}v_x\varphi -v \varphi_x.
 $$
Notice that the last two equations  constitute a Lax pair for the KdV equation (at zero value of the spectral parameter), and the consistency condition $v_{\tau y}=v_{y\tau }$ holds identically modulo the equations for $\varphi$. Setting 
\begin{equation}
w(x, y, \tau)=-\varphi^2, 
\label{w_KdV}
\end{equation}
one can see that $v(x, y, \tau)$ and $w(x, y, \tau)$ solve the system (\ref{max1}). Exact non-singular multi-soliton solutions of these equations can be constructed in the form
\begin{equation}
v=-\frac{d^2}{dx^2}\ln W(\psi_1,\  ..., \psi_n), ~~~ \varphi=\frac{1}{\prod k_i} \frac{W(\psi_1, \  ..., \psi_n, \ 1)}{W(\psi_1, ..., \psi_n)};
\label{v_KdV}
\end{equation}
here $W$ denotes the Wronskian of $n$ functions $\psi_i$ where $\psi_{2j-1} = \cosh(k_{2j-1} x+k_{2j-1}^3y+\tau /k_{2j-1})$ and $\psi_{2j} = \sinh(k_{2j}x+k_{2j}^3y+\tau /k_{2j})$, $k_{i}=const$ satisfying $k_1 < k_2 < ... < k_N$ (e.g., \cite{MS}). Taking $n=2$, 
$k_1=1.05$, $k_2=1.2$, and setting  $\tau=t-y$, we obtain a two-soliton solution of the system (\ref{max}) shown in Fig.~\ref{fig2}. 

  \begin{figure}
[!htbp]
\begin{center}
 \includegraphics[width=60mm]{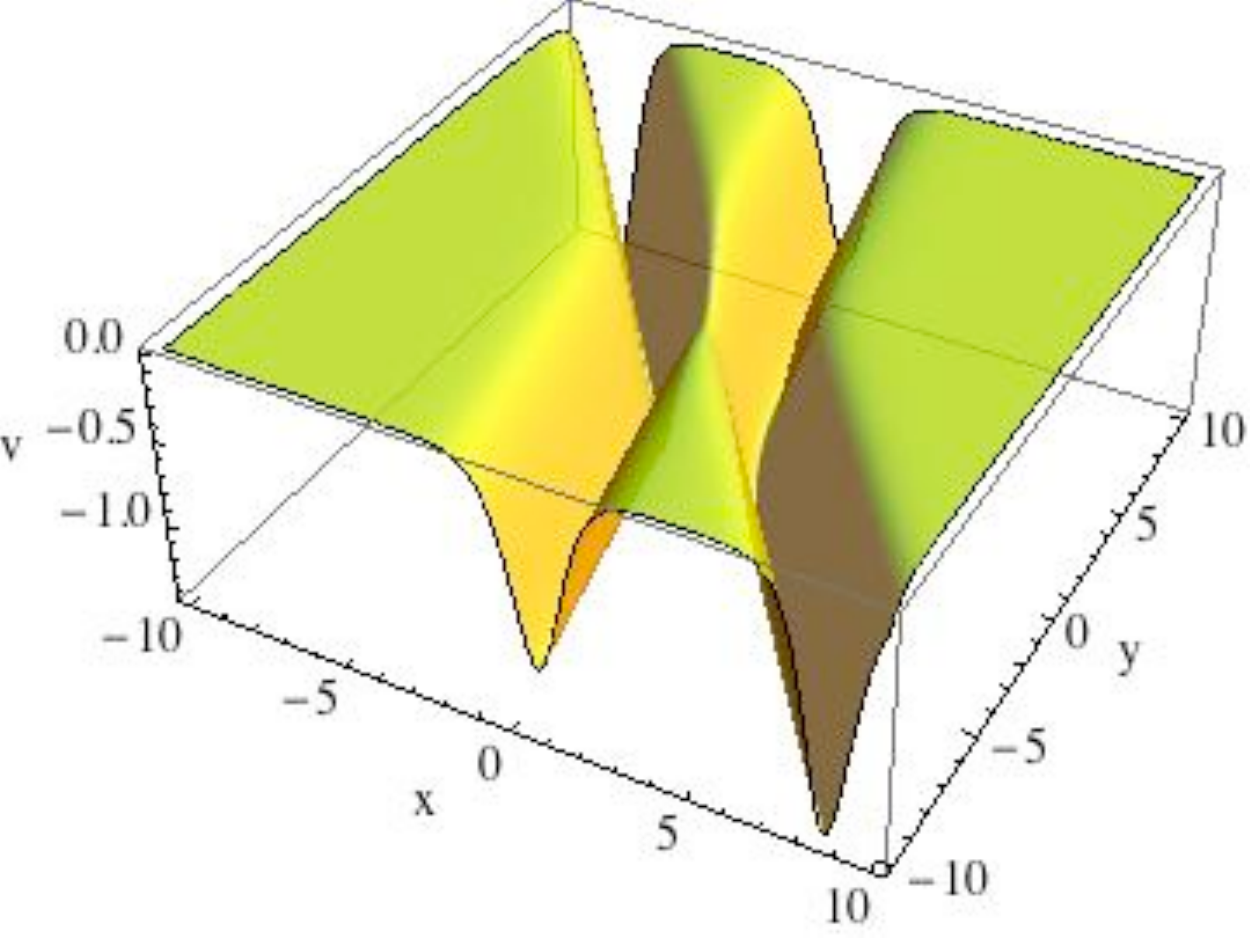} 
 \includegraphics[width=60mm]{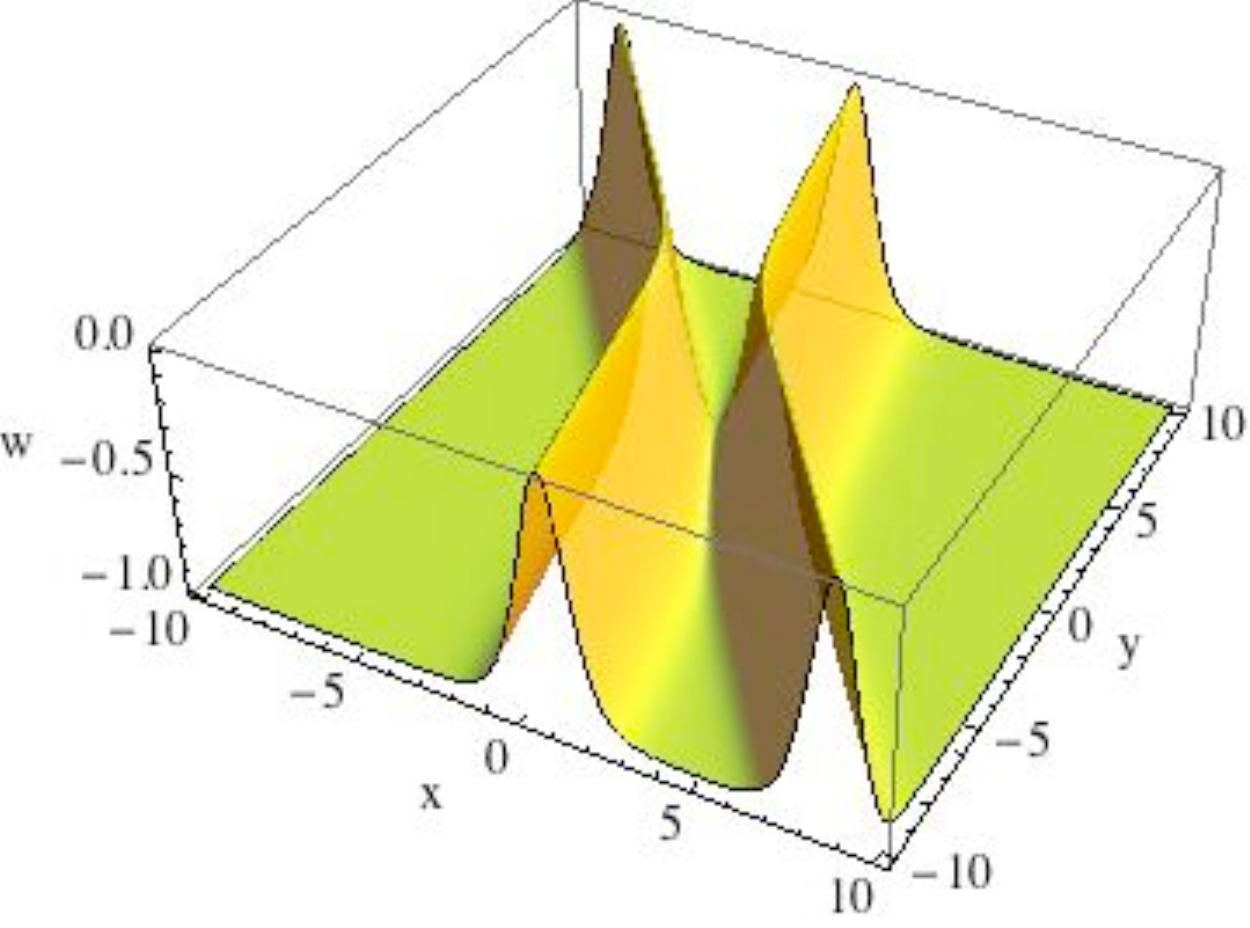} 
\caption{Solution (\ref{w_KdV}), (\ref{v_KdV}) to the system (\ref{max}) for $v$ and $w$  at $t =0$.}
\label{fig2}
\end{center}
\vspace*{-5mm}
\end{figure}

Further differential reductions of the system  (\ref{max}) can be obtained from the general approach of \cite{Shabat1}.

\subsection{Reduction to a pair of (1+1)-dimensional linearly degenerate systems}

The method of hydrodynamic reductions, originally proposed in \cite{Gibb94, GibTsa96} in the context of the dispersionless KP hierarchy, was subsequently generalized to a whole class of multi-dimensional quasilinear systems in  \cite{Fer3, Fer4, Fer5}.   We will use the example of system ({\ref{max})  to illustrate the method, see also  \cite{Pavlov,  Shabat2}. Let us seek $N$-phase solutions of the system (\ref{max})  in the form 
\begin{equation}
v=v(R^1, ..., R^N), ~~~ w=w(R^1, ..., R^N)
\label{an}
\end{equation}
where the phases (Riemann invariants) $R^i(x, y, t)$ satisfy  a pair of compatible $(1+1)$-dimensional systems  
\begin{equation}
R^i_t=\lambda^i(R)R^i_x, ~~~ R^i_y=\mu^i(R)R^i_x.
\label{R}
\end{equation}
Recall that  the commutativity conditions of the flows (\ref{R}) take the form \cite{Tsarev},
\begin{equation}
\frac{\partial_j\lambda
^i}{\lambda^j-\lambda^i}=\frac{\partial_j\mu^i}{\mu^j-\mu^i}, 
\label{comm}
\end{equation}
$i \ne j, ~  \partial_j=\partial/\partial_{ R^j}$, no summation.
The substitution of the ansatz (\ref{an}) into (\ref{max})  implies
$$
\partial_i w=-\lambda^i \partial_iv, ~~~ \mu^i=-\lambda^i-v-\frac{w}{\lambda^i},
$$
which, taking into account
the commutativity condition (\ref{comm}), leads to
$\partial_j\log \lambda^i=\partial_j\log w$,  $\partial_i\partial_j\log w=0, ~\partial_i\partial_jv=0, \ i\ne j$.
These equations are straightforward to integrate:
$$
 v=\sum_k \eta^k(R^k), ~~~ w=\prod_k f^k(R^k),  ~~~ \lambda^i=\varphi^i(R^i)\prod_k f^k(R^k), \\
$$
here $f^i(R^i), \ \eta^i(R^i)$ and $\varphi^i(R^i)$ are arbitrary functions of one variable such that 
$(\eta^i) '=-(f^i)'/(f^i\varphi^i)$. 
After a reparametrization $R^i\to f^i(R^i)$ this simplifies to  
$$
 v=\sum_k \eta^k(R^k), ~~~ w=\prod_k R^k, ~~~  \lambda^i=\varphi^i(R^i)\prod_k R^k, ~~~ (\eta^i) '=-1/(R^i\varphi^i).
$$
From now on we fix a particular choice $\varphi^i(R^i)=1/R^i$ under which both systems in (\ref{R}) become linearly degenerate:
\begin{equation}
R^i_t=(\prod_{k\ne i} R^k)\ R^i_x, ~~~~~ R^i_y=(\sum_{k\ne i} R^k-\prod_{k\ne i} R^k)\ R^i_x.
\label{ld}
\end{equation}
Once a solution to  equations (\ref{ld}) is known, the variables $v$ and $w$ can be reconstructed by the formulae
$$
v=-\sum_k R^k, ~~~ w=\prod_k R^k.
$$
Thus, the $(2+1)$-dimensional system (\ref{max}) is decoupled into a pair of $(1+1)$-dimensional linearly degenerate systems (\ref{ld}) with an arbitrarily large number of dependent variables $N$. Since  the gradient catastrophe for one-dimensional  linearly degenerate hyperbolic systems does not occur  \cite{R1, R2, Liu, Serre}, generic solutions arising from this construction will not break down in finite time. 
The general solution of Eqs. (\ref{ld}) can be represented in the following parametric form,
 \begin{equation}
\begin{array}{c}
\stackrel{R^1} \int \frac{\xi^{N-1}d\xi}{f_1(\xi)}+ \dots+\stackrel{R^N} \int \frac{\xi^{N-1}d\xi}{f_N(\xi)}=x, \\
\ \\
\stackrel{R^1} \int \frac{\xi^{N-2}d\xi}{f_1(\xi)}+ \dots+\stackrel{R^N} \int \frac{\xi^{N-2}d\xi}{f_N(\xi)}=-y, \\
\ \\
\stackrel{R^1} \int \frac{\xi^kd\xi}{f_1(\xi)}+ \dots+\stackrel{R^N} \int \frac{\xi^kd\xi}{f_N(\xi)}=0, ~~~ k=1, 2, ..., N-3,\\
\ \\
\stackrel{R^1} \int \frac{d\xi}{f_1(\xi)}+ \dots+\stackrel{R^N} \int \frac{d\xi}{f_N(\xi)}=(-1)^N(y-t), \end{array}
\label{intn}
\end{equation}
see \cite{Fer}, where $f_i(\xi)$ are $N$ arbitrary functions of one variable. In the next subsections we utilize this representation to construct particular solutions of system (\ref{max}) representing various multiple-wave interactions.

\subsubsection{Reduction to the Jacobi inversion problem}

With the choice $f_i(\xi)=\sqrt{P_{2N+1}(\xi)}$, $i=1,\ldots,N,$ where 
$P_{2N+1}(\xi)$ is a polynomial of degree $2N+1$ in $\xi$, Eqs.
(\ref{intn}) reduce to the Jacobi inversion problem which can solved 
using multi-dimensional theta functions. 

The equation $\mu^{2}=\xi\prod_{i=1}^{2N}(\xi-\lambda_{i})$ with 
$\lambda_{i}\in\mathbb{R}$, $i=1,\ldots,2N$, $\lambda_{i}\neq 
\lambda_{j}$ for $i\neq j$ and $\lambda_{i}\neq0$ 
defines a hyperelliptic Riemann surface 
$\mathcal{L}$ of 
genus $N$ with simple branch points at the $\lambda_{i}$, 0 and at 
infinity. In addition we choose $\sum_{i=1}^{N}\lambda_{i}=0$\footnote{More general hyperelliptic Riemann surfaces can be 
used in this context, but since we only want to illustrate this 
approach, we took this choice leading to simple formulae. With this 
choice, $v$ will be a solution to the KdV equation}. We order the branch points in a way that 
$\lambda_{1}<\lambda_{2}<\ldots<0<\ldots<\lambda_{2N}$. The differentials
$$\eta_{k}=\frac{\xi^{k-1}d\xi}{\mu},\quad k = 1,\ldots,N,$$
form a basis for the holomorphic one-forms on $\mathcal{L}$. On this surface we 
introduce a basis of the homology, where the $a$-cycles are in the 
upper sheet each of them encircling exactly one of the cuts between 
$[\lambda_{1},\lambda_{2}]$, 
$[\lambda_{3},\lambda_{4}]$,\ldots,$[\lambda_{2N-2},\lambda_{2N-1}]$; 
all $b$-cycles start at the cut $[\lambda_{2N},\infty]$, and each of 
them intersects 
exactly one $a$-cycle such that $a_{i}\cdot b_{j}=\delta_{ij}$. A 
normalized basis of holomorphic differentials $\omega_{i}$, 
$i=1,\ldots,N-1$ is given by
$$\omega_{i}=\sum_{j=1}^{N}c_{ij}\eta_{j},\quad i=1,\ldots,N,$$
where the $c_{ij}$ are fixed by the normalization condition
$$\int_{a_{i}}^{}\omega_{j}=2\pi \mathrm{i} \delta_{ij},\quad i,j=1,\ldots,N.$$
With this normalization the matrix 
$$\mathbb{B}_{ij}=\int_{b_{i}}^{}\omega_{j},\quad i,j=1,\ldots,N,$$
is a Riemann matrix, i.e., it is symmetric and has a negative 
definite real part (see for instance \cite{algebro}). Therefore the 
theta function with half integer characteristics $[\alpha] = 
[\alpha';\alpha'']\in \mathbb{R}^{2N}$ defined by
$$\Theta[\alpha](\mathbf{x}|\mathbb{B})=\sum_{\mathbf{n}\in\mathbb{Z}^{N}}^{}
\exp\left(\frac{1}{2}\langle  \mathbf{n+\alpha'},\mathbb{B}(\mathbf{n+\alpha'})\rangle+
\langle \mathbf{n+\alpha'},\mathbf{x}+2\pi \mathrm{i} \alpha''\rangle\right),$$
where the angular brackets denote the Euclidean scalar product, 
is an entire function for all $\mathbf{z}\in \mathbb{C}^{N}.$

With this setting the Jacobi inversion problem following from 
(\ref{intn}) can be written in the form
$$\sum_{j=1}^{N}\int_{}^{R^{j}}\omega_{i}=c_{iN}x-c_{i(N-1)}y+c_{i1}(-1)^{N}(y-t)=:z_{i}.$$
Notice that an arbitrary constant can be added to each $z_{i}$ 
reflecting the freedom to choose a lower limit of the integrals on the 
left-hand side.
Formulae for the symmetric functions $\sum_{j=1}^{N}R^{j}$ and 
$\prod_{j=1}^{N}R^{j}$ in terms of theta functions can be found for 
instance in \cite{algebro}: the former appears in the hyperelliptic 
solutions to the KdV equation, the latter (as a logarithm) in the 
hyperelliptic solutions of the sine-Gordon equation, which is
related to the sinh-Gordon equation by the transformation $u\to 
\mathrm{i}u$. Thus not surprisingly 
both equations which appear as reductions of the studied systems also 
play a role in the context of hyperelliptic solutions. We have 
\begin{equation}
    v = - \sum_{i=1}^{N}R^{i}=-(c+\partial_{xx}\ln \Theta[\alpha](z)),
    \quad w = 
\prod_{i=1}^{N}R^{i}=d+\partial_{xt}\ln \Theta[\alpha](z),
\label{vwhyper}
\end{equation}
where the arbitrary non-singular constant 
characteristic reflects the possibility to change 
$\mathbf{z}$ by an arbitrary constant, and 
where $c$ and $d$ are constants with respect to the physical 
coordinates,
$$
c = \frac{1}{2\pi \mathrm{i}}\sum_{i=1}^{N}c_{iN}\int_{a_{i}}^{}\frac{\xi^{g}d\xi}{\mu},
\quad d = \partial_{xt}\ln 
\Theta[\alpha^{*}]\left(\int_{0}^{\infty}\omega\right),
$$
where $[\alpha^{*}]$ is an arbitrary odd half integer characteristic.
It can be seen that $v_{t}+w_{x}=0$ by construction. The validity of 
the second equation in (\ref{max1}) is checked  by applying 
 derivatives to the theta functions and computing these 
numerically. Since this relation is not implemented in the code, its 
validity provides a strong test on the quality of the numerics. The 
code typically satisfies this test to the order of machine precision. We 
use the code \cite{lmp} developed for hyperelliptic solutions to 
KdV, Kadomtsev-Petviashvili and other equations. In 
Fig.~\ref{jenyahyper3_t0} one can see a typical genus 3 solution.
\begin{figure}
[!htbp]
\begin{center}
\includegraphics[width=14cm]{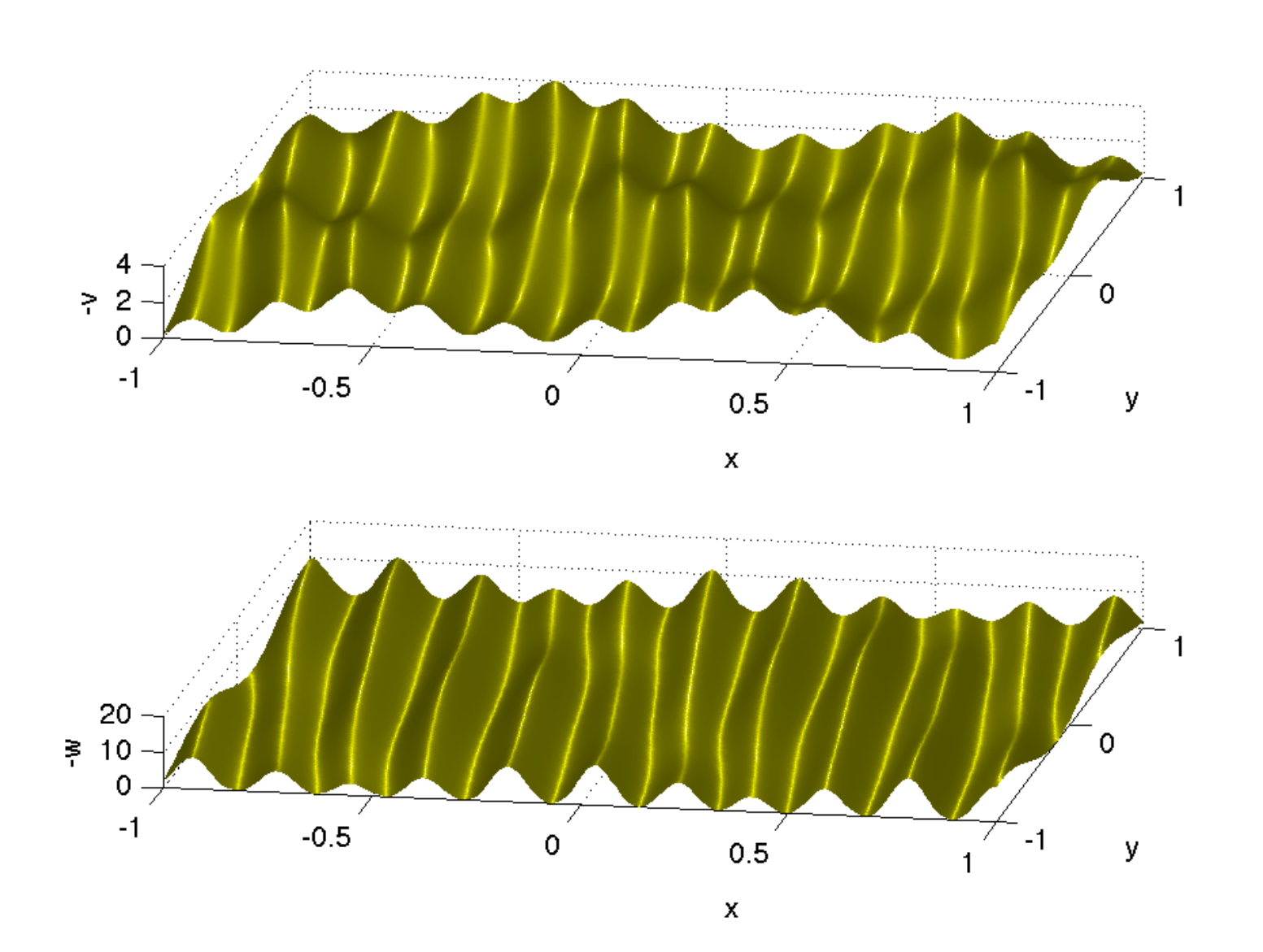} 
\caption{Solutions $v$ and $w$  (\ref{vwhyper}) for $N=3$ on the 
surface with branch points 
$[-3-\epsilon,-3,-\epsilon,0,1,1+2\epsilon,4]$ for $\epsilon=1$
at  $t=0$. 
The characteristics of the theta function are chosen to be
$\alpha_{i}'=1/2$ and 
$\alpha_{i}''=0$ for $i=1,2,3$.}
\label{jenyahyper3_t0}
\end{center}
\vspace*{-5mm}
\end{figure}
As already mentioned, $v$ solves the KdV equation in $y$ and $x$, the 
time $t$ playing just the role of a parameter here. In 
Fig.~\ref{jenyahyper3_t01}, the same situation as in 
Fig.~\ref{jenyahyper3_t0} can be seen for $t=0.1$ thus illustrating 
the time evolution of the pattern.
\begin{figure}
[!htbp]
\begin{center}
\includegraphics[width=14cm]{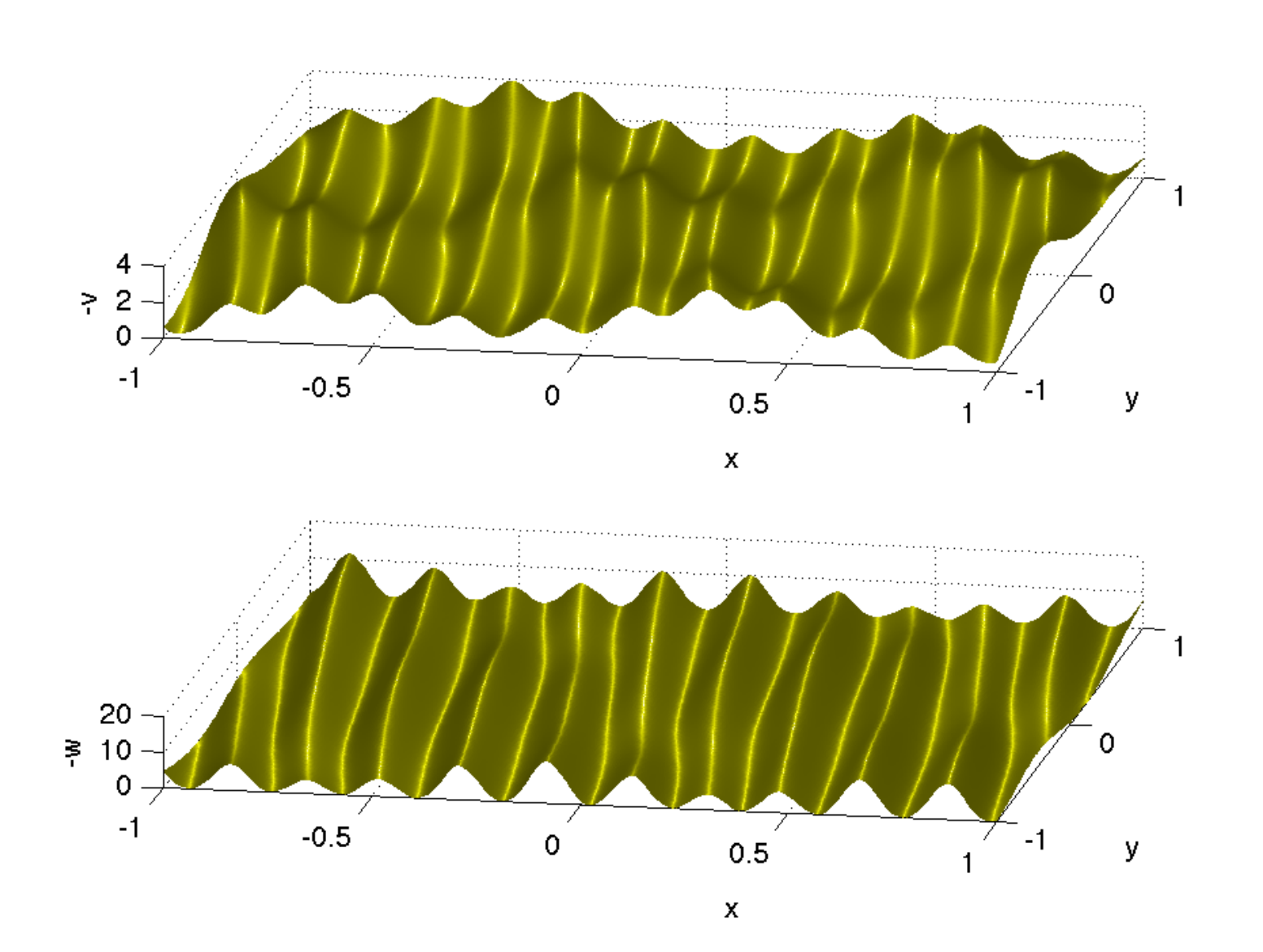} 
\caption{Solutions $v$ and $w$  (\ref{vwhyper}) for $N=3$ on the 
surface with branch points 
$[-3-\epsilon,-3,-\epsilon,0,1,1+2\epsilon,4]$ for $\epsilon=1$
at  $t=0.1$. 
The characteristics of the theta function are chosen to be
$\alpha_{i}'=1/2$ and 
$\alpha_{i}''=0$ for $i=1,2,3$.}
\label{jenyahyper3_t01}
\end{center}
\vspace*{-5mm}
\end{figure}
If the $\epsilon$ in Fig.~\ref{jenyahyper3_t0} tends to 0, the 
surface degenerates which leads to the solitonic limit of the KdV 
solution $v$. This can be seen in Fig.~\ref{jenyahyper3_t0s} where 
a 3-soliton solution to KdV appears. The characteristic phase shift 
at the point of collision between the solitons can be clearly 
recognized. 
\begin{figure}
[!htbp]
\begin{center}
\includegraphics[width=14cm]{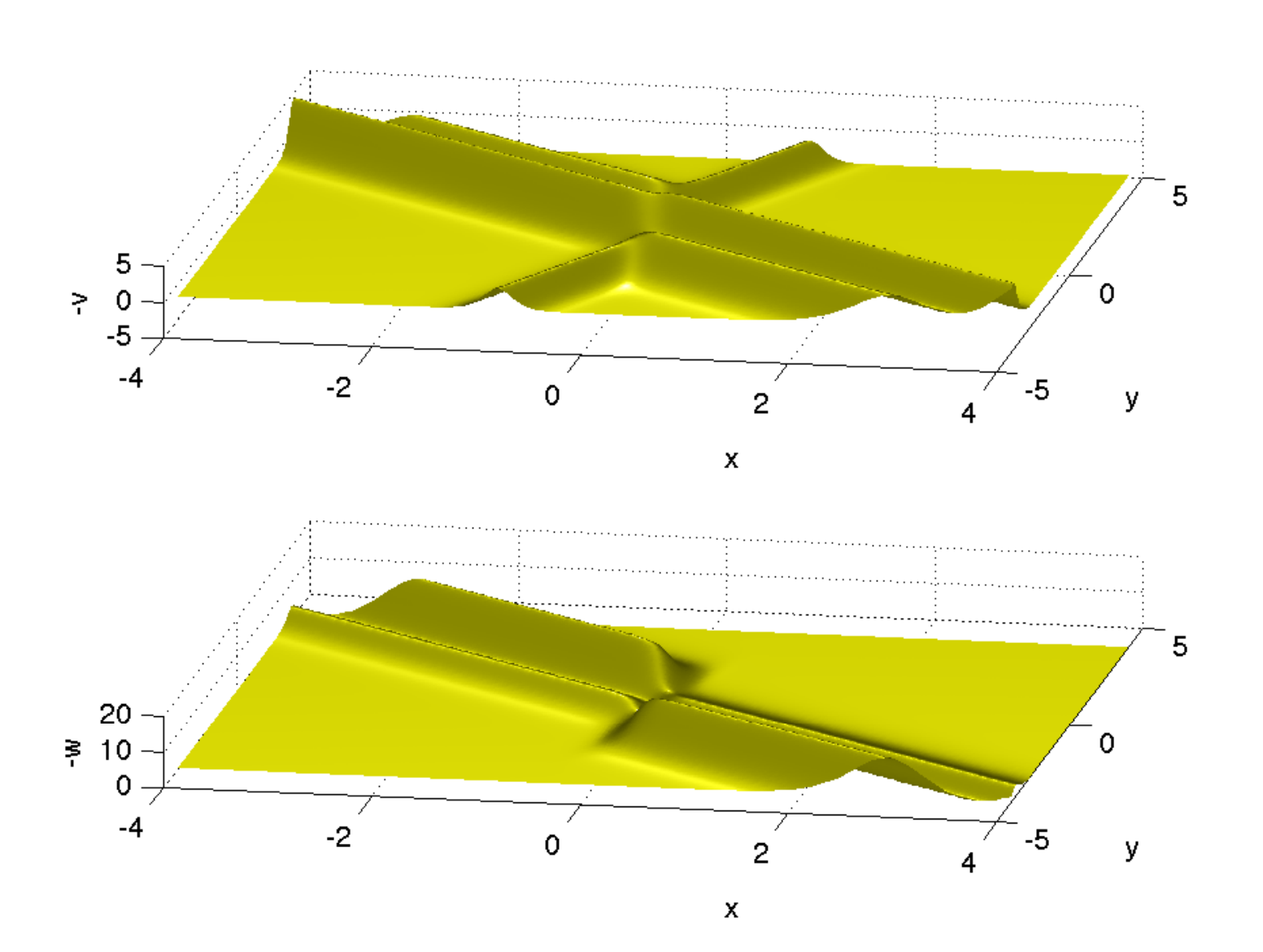} 
\caption{Solutions $v$ and $w$  (\ref{vwhyper}) for $N=3$ on the 
surface with branch points 
$[-3-\epsilon,-3,-\epsilon,0,1,1+2\epsilon,4]$ for $\epsilon=10^{-14}$
at  $t=0$. 
The characteristics of the theta function are chosen to be
$\alpha_{i}'=1/2$ and 
$\alpha_{i}''=0$ for $i=1,2,3$.}
\label{jenyahyper3_t0s}
\end{center}
\vspace*{-5mm}
\end{figure}
In this sense the solutions on a non-degenerate surface can be 
loosely interpreted as an infinite train of 3-solitons. Going to 
higher genus, one obtains $N$-phase solutions which show with 
increasing $N$ more and more structure, as can be seen for instance in 
Fig.~\ref{jenyahyper4_t0}.
\begin{figure}
[!htbp]
\begin{center}
\includegraphics[width=14cm]{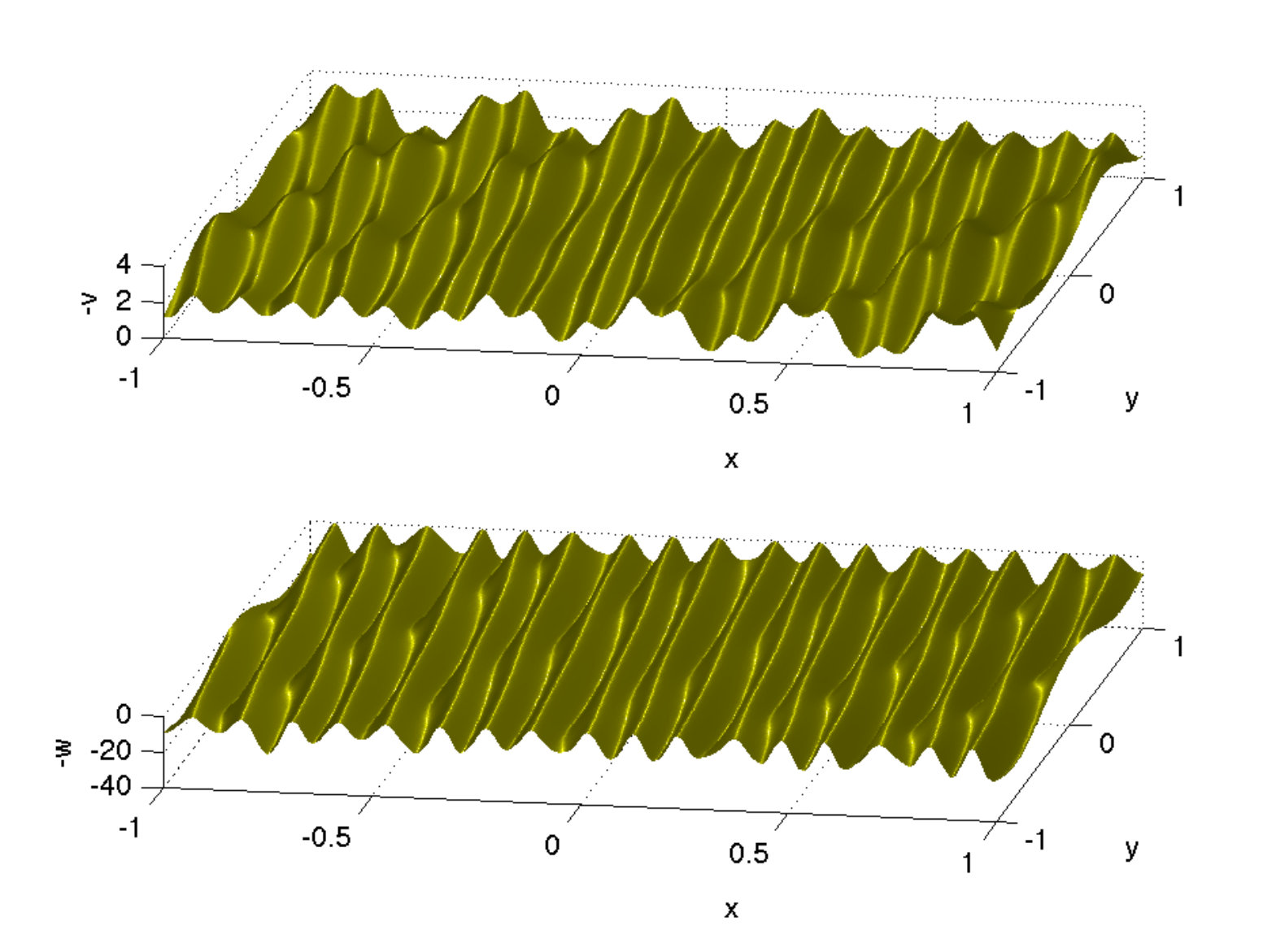} 
\caption{Solutions $v$ and $w$  (\ref{vwhyper}) for $N=4$ on the 
surface with branch points 
$[-4,-3,-2,-1,0,1,2,3,4]$ 
at  $t=0$. 
The characteristics of the theta function are chosen to be
$\alpha_{i}'=1/2$ and 
$\alpha_{i}''=0$ for $i=1,2,3,4$.}
\label{jenyahyper4_t0}
\end{center}
\vspace*{-5mm}
\end{figure}

\subsubsection{Parametric plots of 3-wave interactions}

In the case $N=3$ Eqs. (\ref{ld}) reduce to
\begin{equation}
R^1_t=R^2R^3\ R^1_x, ~~~ R^2_t=R^1R^3\ R^2_x, ~~~ R^3_t=R^1R^2\ R^3_x
\label{e1}
\end{equation}
and
\begin{equation}
R^1_y=(R^2+R^3-R^2R^3)\ R^1_x, ~~~ R^2_y=(R^1+R^3-R^1R^3)\ R^2_x, ~~~ R^3_y=(R^1+R^2-R^1R^2)\ R^3_x,
\label{e2}
\end{equation}
respectively. The formulae (\ref{intn}) simplify to
\begin{equation}
\begin{array}{c}
\stackrel{R^1} \int \frac{\xi^2d\xi}{f(\xi)}+ \stackrel{R^2} \int \frac{\xi^2d\xi}{g(\xi)}+\stackrel{R^3} \int \frac{\xi^2d\xi}{h(\xi)}=x, \\
\ \\
\stackrel{R^1} \int \frac{\xi d\xi}{f(\xi)}+ \stackrel{R^2} \int \frac{\xi d\xi}{g(\xi)}+\stackrel{R^3} \int \frac{\xi d\xi}{h(\xi)}=-y, \\
\ \\
\stackrel{R^1} \int \frac{d\xi}{f(\xi)}+ \stackrel{R^2} \int \frac{d\xi}{g(\xi)}+\stackrel{R^3} \int \frac{d\xi}{h(\xi)}=t-y, \end{array}
\label{int}
\end{equation}
where $f, g, h$ are arbitrary functions. 
For our purposes, it will be more convenient to work with an equivalent  parametric representation of the general solution of (\ref{e1})--(\ref{e2}), namely,
$$
R^1=F(a), ~~~ R^2=G(b), ~~~ R^3=H(c),
$$
where $a, b, c$ are functions of $x, y, t$ defined implicitly via
\begin{equation}
\begin{array}{c}
\int \frac{F^2(a)}{F(a)-1} da+\int \frac{G^2(b)}{G(b)-1} db+\int \frac{H^2(c)}{H(c)-1} dc=x, \\
\ \\
\int \frac{F(a)}{F(a)-1} da+\int \frac{G(b)}{G(b)-1} db+\int \frac{H(c)}{H(c)-1} dc=-y, \\
\ \\
a+b+c=-t; 
\end{array}
\label{e44}
\end{equation}
here $F, G, H$ are three arbitrary functions, and all integrals are understood as indefinite. The corresponding solution of (\ref{max}) is given by the formulae
 \begin{equation}
 v=-F(a) -G(b) -H(c),  ~~~ w=F(a) G(b) H(c).
 \label{vw}
 \end{equation}
Relations (\ref{e44}) can be derived from (\ref{int}) by setting
$$
R^1=F(a), ~~ R^2=G(b), ~~ R^3=H(c)
$$
and
$$
\stackrel{R^1} \int \frac{(\xi-1) d\xi}{f(\xi)} = a, ~~~ \stackrel{R^2} \int \frac{(\xi-1) d\xi}{g(\xi)} = b, ~~~ \stackrel{R^3} \int \frac{(\xi-1)d\xi}{h(\xi)}= c,
$$
respectively.  Let us choose, for instance, three soliton-like profiles
 \begin{equation}
 F(a)=\alpha- \frac{\epsilon}{\cosh^2\mu a}, ~~~  G(b)=\beta- \frac{\epsilon}{\cosh^2\mu b}, ~~~ H(c)=\gamma- \frac{\epsilon}{\cosh^2\mu c},
 \label{FGH}
 \end{equation}
 where $\alpha, \beta, \gamma$ are distinct constants 
 and $\epsilon$ is a small parameter. In this case equations (\ref{e44}) take the form
 $$
 \frac{\alpha^2a}{\alpha-1} +\frac{\beta^2b}{\beta-1} +\frac{\gamma^2c}{\gamma-1} +\epsilon (...)=x, ~~~  \frac{\alpha a}{\alpha-1} +\frac{\beta b}{\beta-1} +\frac{\gamma c}{\gamma-1} +\epsilon (...)=-y, ~~~  a+b+c=-t,
 $$
 where dots denote certain explicit functions of $a, b, c$ which are bounded along with their first order derivatives. Thus, for sufficiently small $\epsilon$ the change of variables $a, b, c \to x, y, t$ is nonsingular and well-defined globally, so that the corresponding $v$ and $w$ will  also be nonsingular. 
Expressing $c$  in the form $c=-t-a-b$ and substituting  into the formulas (\ref{vw}) for $v$, $w$, and the first two equations  (\ref{e44}), one obtains $v, w, x, y$ as functions of $a$ and $b$. Thus, one can construct parametric plots of $v(x, y, t)$ and $w(x, y, t)$. 
 Figs.  \ref{fig3} and  \ref{fig4} show snapshots of $v$ and $w$   at $t=0$ and $t=6$ 
for the parameter values $\alpha =1 - \sqrt {\frac{(1 + \tan (\phi + \pi/3) (1 + \tan (\phi - \pi/3))}{ 1 + \tan \phi}}, \  \beta =1 - \sqrt {\frac{(1 + \tan \phi) (1 + \tan (\phi - \pi/3))}{1 + \tan (\phi + \pi/3)}}, 
\ \gamma =1 - \sqrt {\frac{(1 + \tan \phi) (1 + \tan (\phi + \pi/3))}{ 1 + \tan (\phi - \pi/3)}}$,
where $\phi = \pi/9$, and $\mu = 1,\ \epsilon=0.05$. At $t=0$ initial profiles represent three solitons having a triple collision at the origin (parameters are chosen in such a way that 
 solitons meet at equal angles $\pi/3$).  The triple collision then disintegrates  into  pairwise interactions, shown at $t=6$.

\begin{figure}
[!htbp]
\begin{center}
\includegraphics[width=70mm]{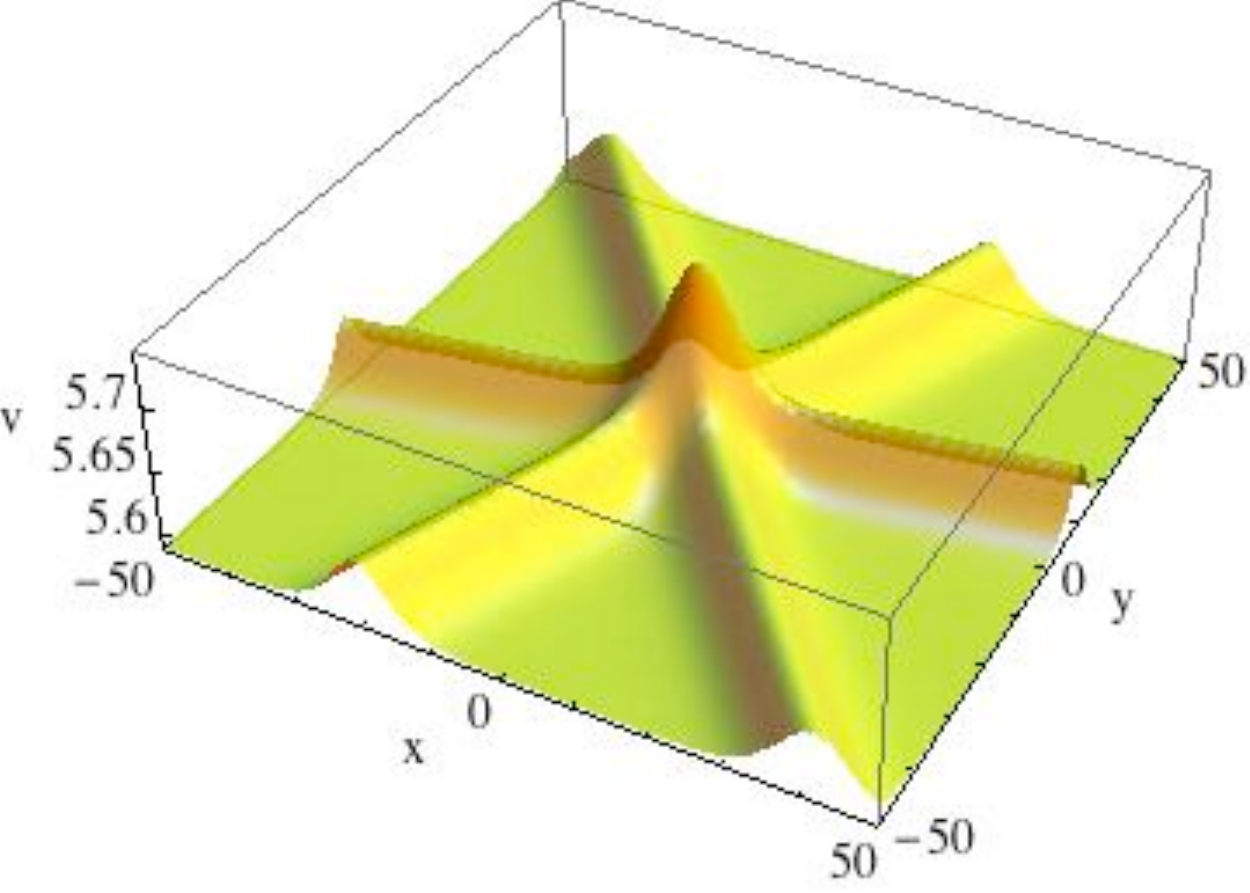} 
\includegraphics[width=75mm]{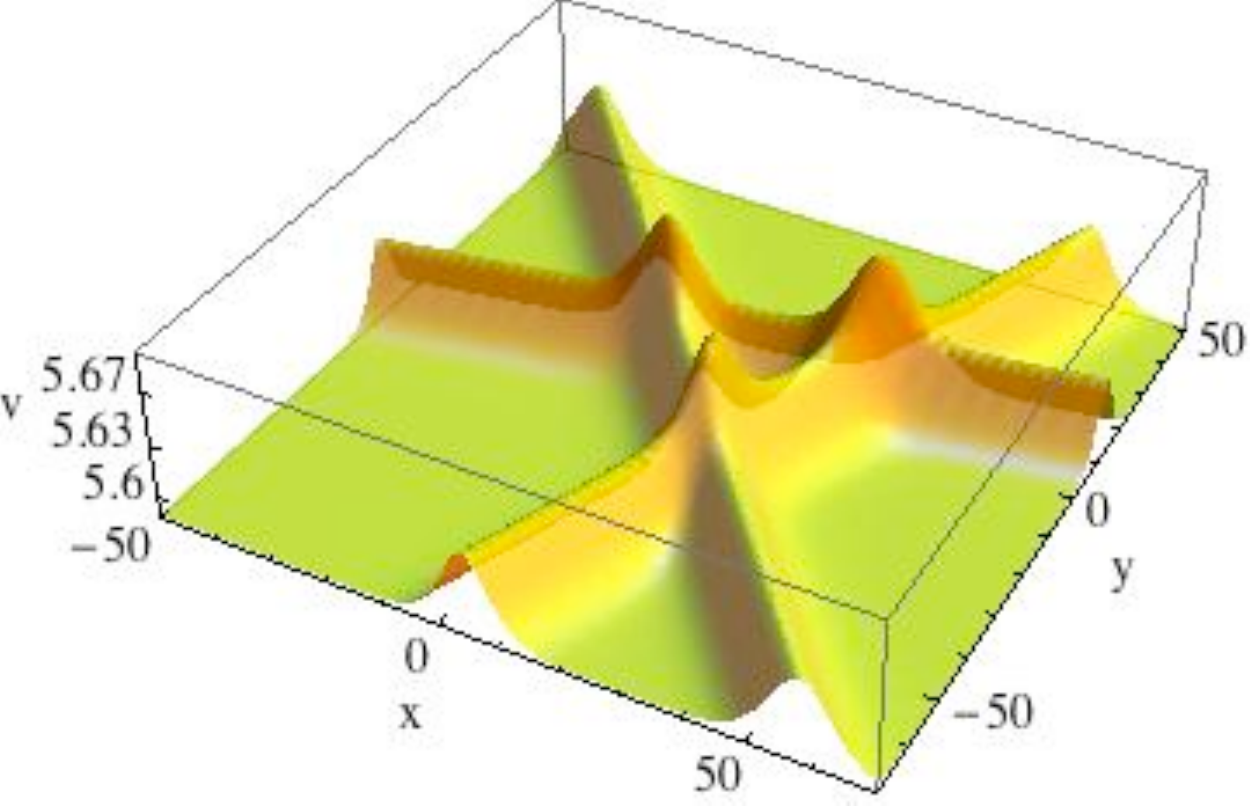}
\caption{Solution (\ref{vw}), (\ref{FGH}) to the system (\ref{max}) for $v$  at $t=0$ and $t=6$.}
\label{fig3}
\end{center}
\vspace*{-5mm}
\end{figure}

\begin{figure}
[!htbp]
\begin{center}
\includegraphics[width=70mm]{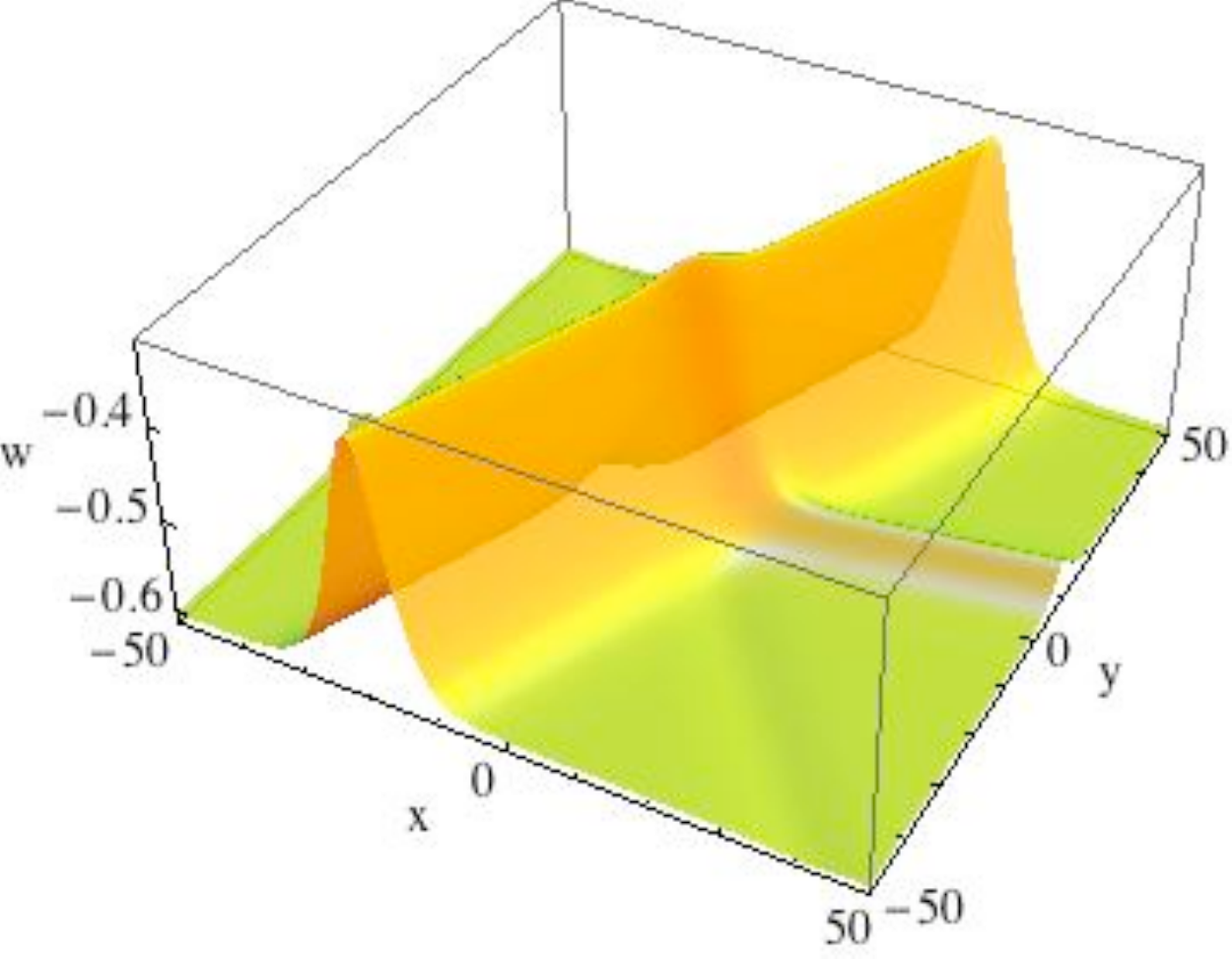} 
\includegraphics[width=75mm]{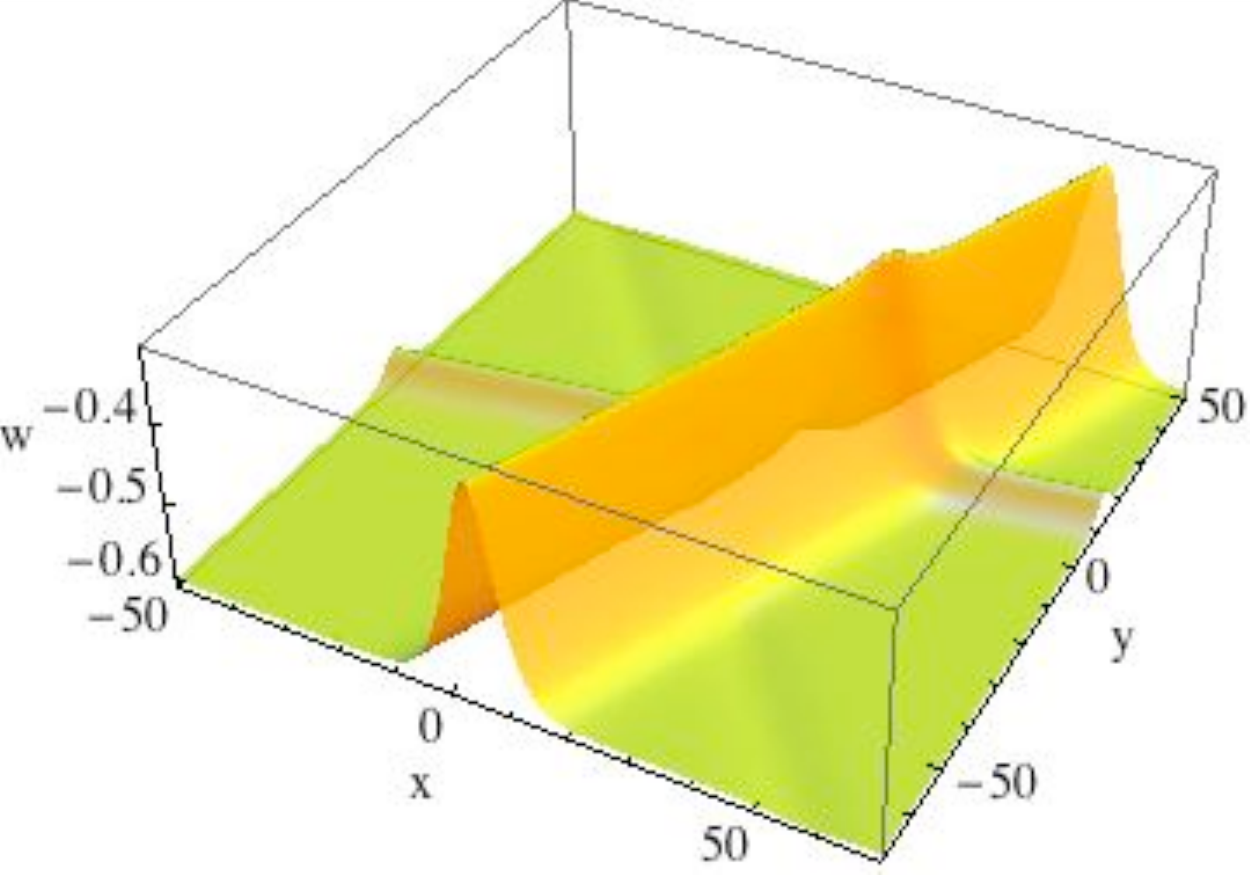}
\caption{Solution (\ref{vw}), (\ref{FGH}) to the system (\ref{max}) for $w$  at $t=0$ and $t=6$.}
\label{fig4}
\end{center}
\vspace*{-5mm}
\end{figure}
Unfortunately, for $N>3$ one cannot  reduce Eqs. (\ref{intn}) to parametric formulas, and has to solve them numerically. 

\subsubsection{Plots of $N$-wave interactions}

To obtain a more convenient form for explicit integration, 
we rewrite Eqs. (\ref{intn}) as
 \begin{equation}
\begin{array}{c}
\int F_1^{N-1}(p_1) dp_1+\dots+\int F_N^{N-1}(p_N) dp_N=x, \\
\ \\
\int F_1^{N-2}(p_1) dp_1+\dots+\int F_N^{N-2}(p_N) dp_N=-y, \\
\ \\
\int F_1^{k}(p_1) dp_1+\dots+\int F_N^{k}(p_N) dp_N=0,  ~~~ k=1, 2, ..., N-3,\\
\ \\
p_1+... +p_N=(-1)^N(y-t).
\end{array}
\label{en}
\end{equation}
here $F_i(p_i)$ are  arbitrary functions, and all integrals are understood as indefinite. The corresponding solution of (\ref{max}) is given by the formulae
$$
 v=-\sum F_i(p_i) ,  ~~~ w=\prod F_i(p_i).
$$
Formulae  (\ref{intn}) and (\ref{en}) are identified upon setting $R^i=F_i(p_i), \  \stackrel{R^i} \int \frac{d\xi}{f_i(\xi)}=p_i$.
Equations   (\ref{en}) can  be used to construct nonlinear interactions of $N$
soliton-like profiles  by taking, e.g., 
$ F_i(p_i)=\alpha_i + \frac{\epsilon}{\cosh^2\mu_{i} p_i}$,  where $\epsilon$ is sufficiently small and the constants $\alpha_i$ are chosen in such a way that  equations  (\ref{en}) are uniquely solvable for $p_i(x, y, t)$.  

This approach can in principle be used to build hump-like initial 
data by letting $N$ tend to infinity, thus providing in a loose sense a non-linear 
generalization of Fourier analysis for the studied PDE. We illustrate 
this concept below by studying examples for finite $N$ numerically. 
The idea is to construct for $t=0$ a central hump from $N$ elementary 
solutions of the form of the KdV soliton. 

The angles between these elementary solutions are chosen to be equal. 
From the second relation in (\ref{ld}) it can be seen that a 
condition on these 
angles is equivalent to a condition on the $\alpha_{i}$,
\begin{equation}
    \sum_{j=1}^{N}\alpha_{j} -\alpha_{i} - 
    \frac{1}{\alpha_{i}}\prod_{j=1}^{N}\alpha_{j} = \tan \psi_{i}, 
    \quad  i = 1,\ldots,N
    \label{acon}.
\end{equation}
To ensure that $w(x,y,0)<0$ we choose 
$F_{i}=\alpha_{i}+\frac{\epsilon}{\cosh^{2}\mu_{i}p_{i}}$ for 
$i=1,\ldots,N-2$ and $F_{i}=\alpha_{i}$ for $i=N-1,N$. In addition we 
fix $\sum_{j=1}^{N}\alpha_{j}=0$ and $\prod_{j=1}^{N}\alpha_{j}=-1$, 
thus assuring that we are studying solutions on the zero background for 
$v$ and on the $-1$ background for $w$. The first $N-2$ $\alpha_{i}$ then follow 
from (\ref{acon}) via the quadratic relations,
$$1/\alpha_{i}-\alpha_{i}=\tan \psi_{i},\quad i=1,\ldots,N-2.$$
The remaining two $\alpha_{i}$ are a consequence of the condition of vanishing 
sum of the $\alpha_{i}$ and product equal to $-1$,
$$\alpha_{N-1,N}= \frac{1}{2}\left(\sum_{j=1}^{N-2}\alpha_{j} \pm
\sqrt{\left(\sum_{j=1}^{N-2}\alpha_{j}\right)^2-\frac{4}{\prod_{j=1}^{N-2}\alpha_{j}}}\right).$$
The $\psi_{i}$ are chosen in a way to generate equal angles between 
the elementary solutions in the $(x,y)-$plane for $t=0$. For 
numerical reasons one has to avoid $\psi=\pi/2$ since the tangent 
diverges there. To obtain finite $\psi_{i}$ also for even $N$, we 
shift them by a common factor, thus
$$\psi_{n} = 
\begin{cases}
    \frac{n\pi}{N-2} & N \mbox{ odd,}  \\
    \frac{n\pi}{N-2} + \frac{\pi}{2(N-2)} & N \mbox{ even.}
\end{cases}
$$ 
With this choice we assure that the values for $\tan \psi$ are as 
small as possible. This is numerically important since the 
$\alpha_{i}$ appear in an $N\times N$ Vandermonde determinant, i.e., 
there will be $N$th powers of each $\alpha_{i}$. For the same reason 
we also have to choose the signs carefully to avoid too large or too small 
values of $\alpha_{N-1,N}$. With the above choice of the $\psi_{i}$, 
$i=1,\ldots,N-2 $, we have assured that with some $\alpha_{j}$ a 
solution to one of the equations (\ref{acon}), also $-\alpha_{j}$ and 
$\pm \alpha_{j}$ are solutions to these equations (either for 
$\psi_{j}$ or for $\pi-\psi_{j}$). We fix the signs in a way to 
obtain a minimal $\sum_{i=1}^{N-2}\alpha_{i}$ and a negative 
$\prod_{i=1}^{N-2}\alpha_{i}$. This is done by alternating the signs 
in the $\alpha_{i}$, $i=1,\ldots,N_{max}$ ($N_{max}=N/2-1$ for $N$ 
even and $N_{max}=(N-1)/2$ for $N$ odd). The remaining signs are 
chosen in a way that the $\alpha_{j}$ corresponding to $\pi-\psi_{i}$ 
is equal to $1/\alpha_{i}$. Thus we assure that $\alpha_{N-1}$ and 
$\alpha_{N}$ are real and that $1/|\alpha_{N_{max}}|\leq 
|\alpha_{i}|\leq |\alpha_{N_{max}}|$. We give the values of the 
$\alpha_{i}$ for two examples below. Since $\alpha_{N_{max}}\sim 
2(N-2)/\pi$, we can in double precision reach values of $N=14$ in 
which case the conditioning of the Vandermonde matrix of the 
$\alpha_{i}$ is of the order of $10^{-14}$ (the absolute value of the ratio of the smallest 
to the largest eigenvalue of the matrix). To treat higher values of 
$N$ more than double precision is needed which is, however, beyond 
the scope of this paper.

The goal of this subsection is to construct for $t=0$ a single hump 
for $v$ with value $c=const$ at $x=y=0$. To this end we choose 
$\epsilon=c/(N-2)$ since $N-2$ elementary solutions will overlap 
at the origin of the $(x,y)$ plane with the above choices. Thus we 
take $\epsilon = c/(N-2)$ and take $c=0.1$. The constant $c$ has to 
be chosen in a way that $w$ remains negative which seems to be the 
case in the 
numerical experiments discussed below if $c<1$. A 
smaller value of $c$ leads to a more rapid convergence of the 
iteration to be described below.

For the given form of the $F_{i}$, $i=1, ..., N$, we have ($n\geq 1$)
$$\int_{}^{}F_{i}^{n}(p_{i})dp_{i} = 
\alpha_{i}^np_{i}+\sum_{j=1}^{n}
\begin{pmatrix}
    n  \\
    j
\end{pmatrix}
\epsilon^{j}\alpha_{i}^{n-j}I_{j}(p_{i}),
$$
where $I_{n}(p)=\int_{}^{p}dp'/\cosh^{2n}p'$. 
Equations (\ref{en}) thus
take the form ($ k = 
    1,2,\ldots,N-3$)
\begin{align*}
    \sum_{i=1}^{N-2}\left\{\alpha_{i}^{N-1}p_{i}+\sum_{j=1}^{N-1}
    \begin{pmatrix}
	N-1  \\
	j
    \end{pmatrix}
    \epsilon^{j}\alpha_{i}^{N-1-j}I_{j}(p_{i})\right\}+\alpha^{N-1}_{N-1}p_{N-1}+\alpha_{N}^{N-1}p_{N} & =x,  \\
    \sum_{i=1}^{N-2}\left\{\alpha_{i}^{N-2}p_{i}+\sum_{j=1}^{N-2}
	\begin{pmatrix}
	    N-2  \\
	    j
	\end{pmatrix}
	\epsilon^{j}\alpha_{i}^{N-2-j}I_{j}(p_{i})\right\}+\alpha^{N-2}_{N-1}p_{N-1}+\alpha_{N}^{N-2}p_{N} & =-y  \\   
	\sum_{i=1}^{N-2}\left\{\alpha_{i}^kp_{i}+\sum_{j=1}^{k}
	    \begin{pmatrix}
		k  \\
		j
	    \end{pmatrix}
	    \epsilon^{j}\alpha_{i}^{k-j}I_{j}(p_{i})\right\}+\alpha^{k}_{N-1}p_{N-1}+\alpha_{N}^{k}p_{N} & =0,  \\
    p_{1}+\ldots+p_{N} & =(-1)^{N}(y-t),
\end{align*}
This system can be 
solved iteratively. Let $V$ be the Vandermonde determinant of the 
$\alpha_{i}$. Then we solve iteratively for the vector $\vec{p}$, 
where $\vec{p}_{n+1}$ follows from the $n$th iterate via
$$V\vec{p}_{n+1}=\vec{J}(\vec p_{n}),$$
with 
$$
J_{1}(\vec{p}_{n})=-\sum_{i=1}^{N-2}\sum_{j=1}^{N-1}
    \begin{pmatrix}
	N-1  \\
	j
    \end{pmatrix}
    \epsilon^{j}\alpha^{N-1-j}_{i}I_{j}(p_{i})+x,
$$
etc, and 
$$\vec{J}(\vec{p}_{0})=(x,-y,0,\ldots,0,(-1)^{N}(y-t)).$$
The iteration is stopped once $|\vec{p}_{n+1}-\vec{p}_{n}|<\delta$ 
(we use $\delta=10^{-6}$). For a value of $c\ll 1$, this iteration 
converges rapidly (typically less than 4 iterations). If $c\sim 1$ 
the iteration converges only slowly or not at all.

The results of the above procedure for $N=6$ (four elementary 
solutions) for $t=0$ can be seen in Fig.~\ref{vw6}. Since this 
approach is concerned with hump-like structures in $v$, the outcome 
for $w$ is not obvious, but can be seen in the lower part.
\begin{figure}
[!htbp]
\begin{center}
  \includegraphics[width=14cm]{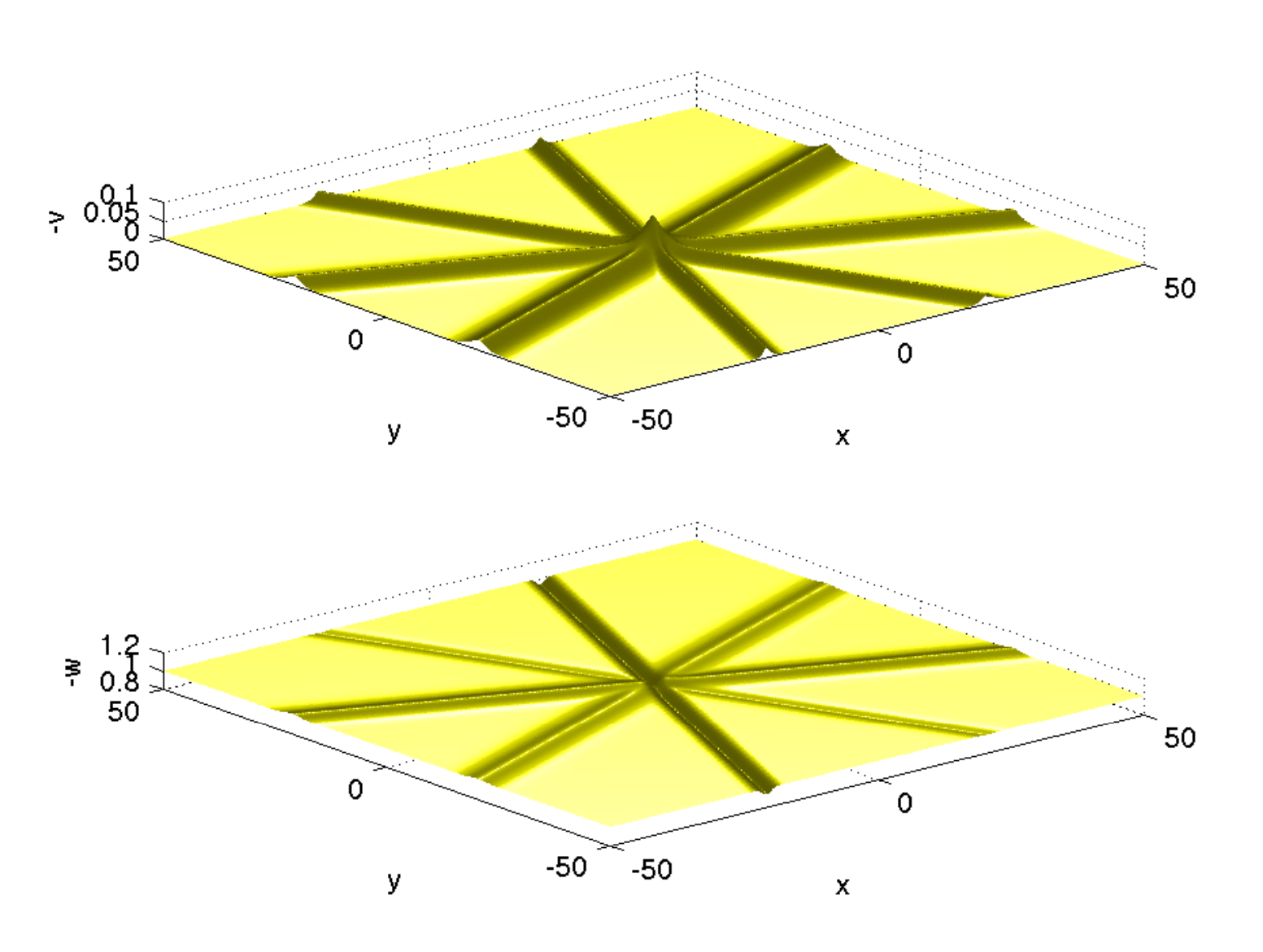} 
\caption{Solutions $v$ and $w$ to the system (\ref{en}) for $N=6$ 
at  $t=0$ and $\vec{\mu}=(1,100,1,1)$. The values of $\vec{\mu}$ have 
been chosen in a way that the lateral extension of the elementary 
waves is of a similar order. This implies $\vec{\alpha}\approx(0.81,   -2.77,   -0.36,    1.23,   -0.59,    1.69)$.}
\label{vw6}
\end{center}
\vspace*{-5mm}
\end{figure}
The central hump for $v$ at the origin of the $(x,y)$-plane 
disintegrates with time as can be seen in Fig.~\ref{vw6t6}.
\begin{figure}
[!htbp]
\begin{center}
  \includegraphics[width=14cm]{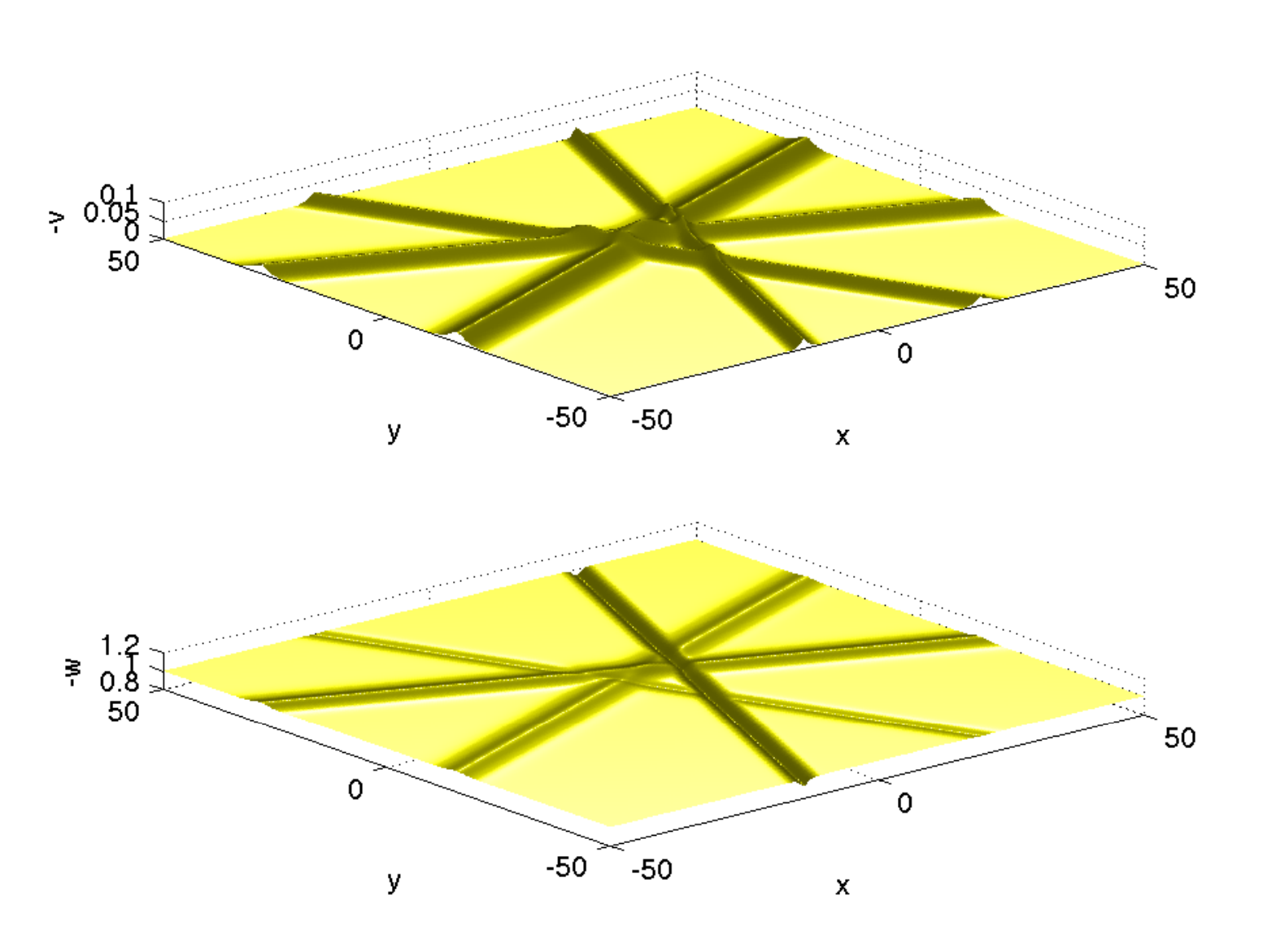} 
\caption{Solutions $v$ and $w$ to the system (\ref{en}) for $N=6$ 
at  $t=6$ and $\vec{\mu}=(1,100,1,1)$.}
\label{vw6t6}
\end{center}
\vspace*{-5mm}
\end{figure}

Since we fixed the maximum of the absolute value of $v$ at the center 
to be of the order of $c$, the relative maximum of the modulus of the 
elementary waves tends with large $N$ to 0. This is illustrated in 
Fig.~\ref{vw13} for $N=13$.
\begin{figure}
[!htbp]
\begin{center}
  \includegraphics[width=14cm]{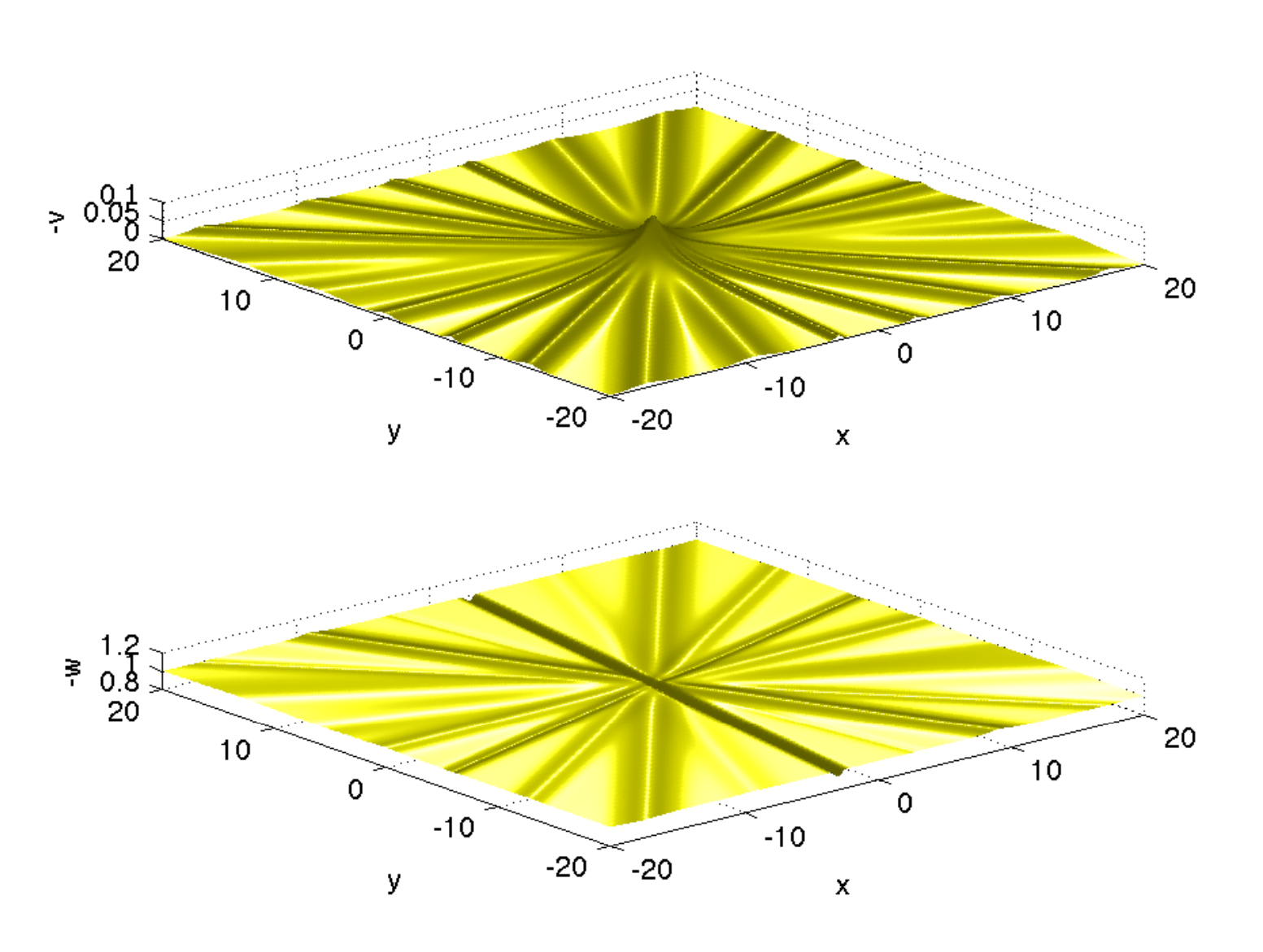} 
\caption{Solutions $v$ and $w$ to the system (\ref{en}) for $N=13$ 
at  $t=0$ and $\vec{\mu}=(100,100,10,10^{4},10, 10^{10}, 1,10^{5} , 
10,10^{3},100)$. The values of $\vec{\mu}$ have 
been chosen in a way that the lateral extension of the elementary 
waves is of a similar order. This implies  $\vec{\alpha}\approx( 1.00,   -1.16,    0.73,   -1.73,   
0.39,   -7.10,   -0.14,  2.58,   -0.58,    1.37,   
-0.86,   -0.18,    5.68)$.}
\label{vw13}
\end{center}
\vspace*{-5mm}
\end{figure}
The central hump for $v$ at the origin of the $(x,y)$-plane 
disintegrates with time as can be seen in Fig.~\ref{vw13t2}. Compare 
this figure with the time evolution of hump-like initial data for $v$ 
studied numerically in the next subsection.
\begin{figure}
[!htbp]
\begin{center}
  \includegraphics[width=14cm]{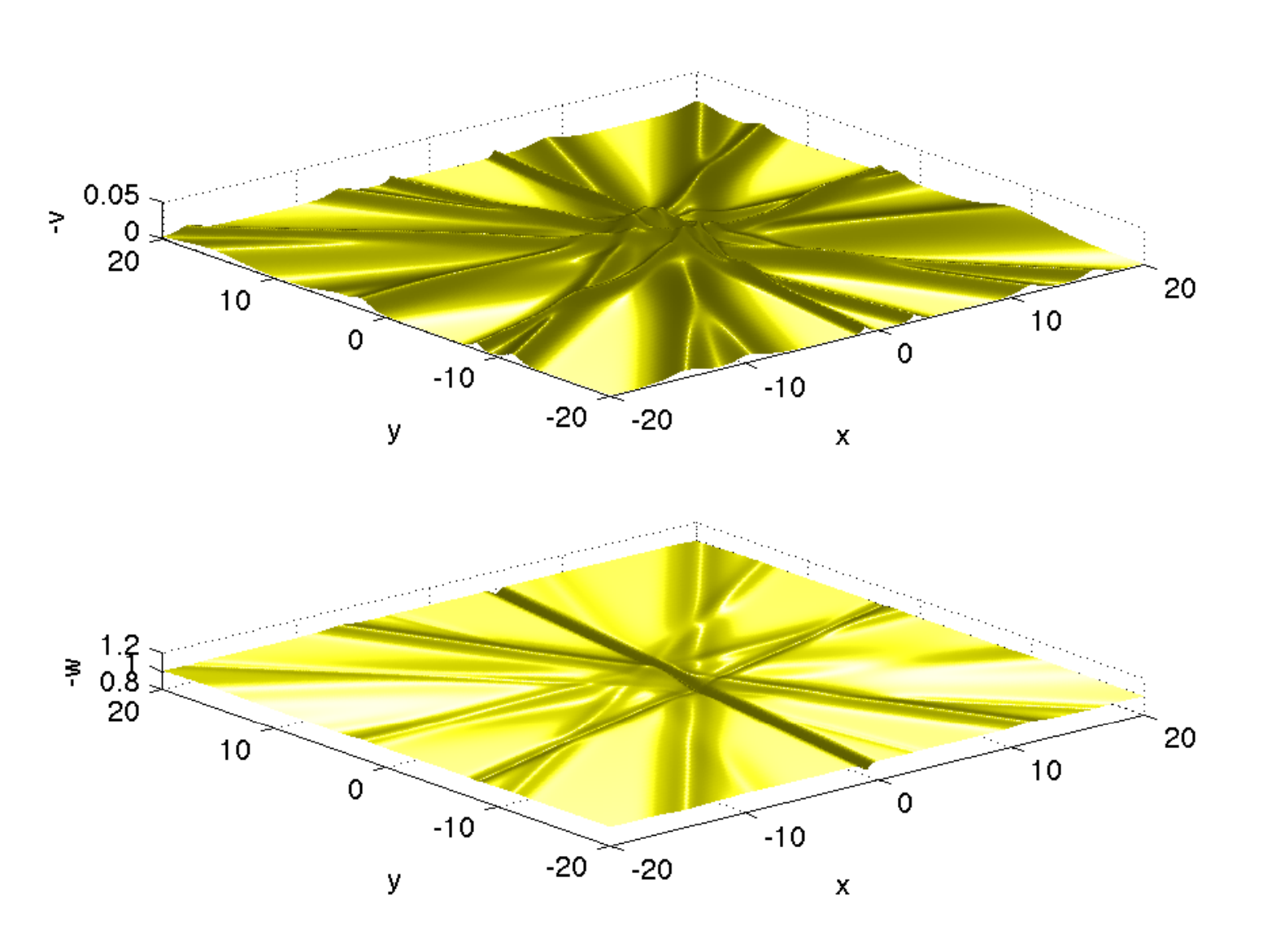} 
\caption{Solutions $v$ and $w$ to the system (\ref{en}) for $N=13$ 
at  $t=2$ and $\vec{\mu}=(100,100,10,10^{4},10, 10^{10}, 1,10^{5} , 
10,10^{3},100)$.}
\label{vw13t2}
\end{center}
\vspace*{-5mm}
\end{figure}

\subsection{Cauchy problems and numerical evaluations}
We will use the Friedrichs-symmetrized form (\ref{max11}),  which is well-posed for  $w< 0$, for numerical simulations of system (\ref{max}). 
The task is to  investigate the conjecture that solutions to the 
system (\ref{max}) for generic initial data do not develop a point of 
gradient catastrophe in finite time. 

To this end we consider initial data in the Schwartzian class of rapidly 
decreasing functions. Such functions can be treated for numerical 
purposes as effectively periodic if the period is chosen large enough 
that the studied functions decrease below the machine precision 
(which is of the order $10^{-16}$ for double precision computations in 
MATLAB \cite{Matlab}) at the ends of the intervals. The Gibbs phenomenon in such a case is of the order of the 
rounding error and thus negligible. This setting allows the use of 
Fourier spectral methods which are known for their excellent 
approximation properties for smooth functions. Thus we will 
approximate the spatial dependence of the functions $v$ and $w$ via 
discrete Fourier series. We find that the solutions to the system 
(\ref{max}) stay in the Schwartzian class in contrast, e.g.,  to 
solutions to the dispersionless Kadomtsev-Petviashvili (dKP) 
equation, see \cite{ksm07} for a numerical study.

The time dependence is treated via a standard fourth order 
Runge-Kutta scheme. The quality of the numerics is controlled via the 
conserved quantity 
$$I:=\int_{-\infty}^{\infty}\int_{-\infty}^{\infty}(v^{2}-(w-w_\infty))dxdy$$
of system (\ref{max}), where $w_\infty=\lim_{x,y\to\infty}w(x,y,t)$. Since the 
conservation of this quantity is not implemented in the code, it will 
not be exactly observed in actual numerical experiments due to 
unavoidable numerical errors. Thus the numerical conservation of $I$
can serve as an indicator for the achieved accuracy. In our experiments the 
error in the relative conservation of this quantity was below 
$10^{-6}$. It is known (see for instance \cite{etna}) that integral quantities overestimate the 
accuracy in a  $L_{\infty}$ sense of numerical solutions to a partial differential equations 
by one to two orders of magnitude. Thus the results presented below 
are accurate to at least $10^{-4}$ which is well beyond plotting 
accuracy. Comparing the behavior of the numerical solutions to 
solutions to the dKP equation in \cite{ksm07} which show blow up, we 
see in the experiments performed below no indication of an increasing 
gradient of the solutions. Thus it seems that at least for the class of initial data 
considered, the numerical solutions will have global existence in time.

As a concrete example we study the initial data 
$v(x,y,0)=\exp(-(x^{2}+y^{2}))$ and $w(x,y,0)=-1$. It can be seen in 
Fig.~\ref{vgauss_4td} that the solution for $v$ stays smooth and localized in 
both spatial directions. 
\begin{figure}
[!htbp]
\begin{center}
  \includegraphics[width=14cm]{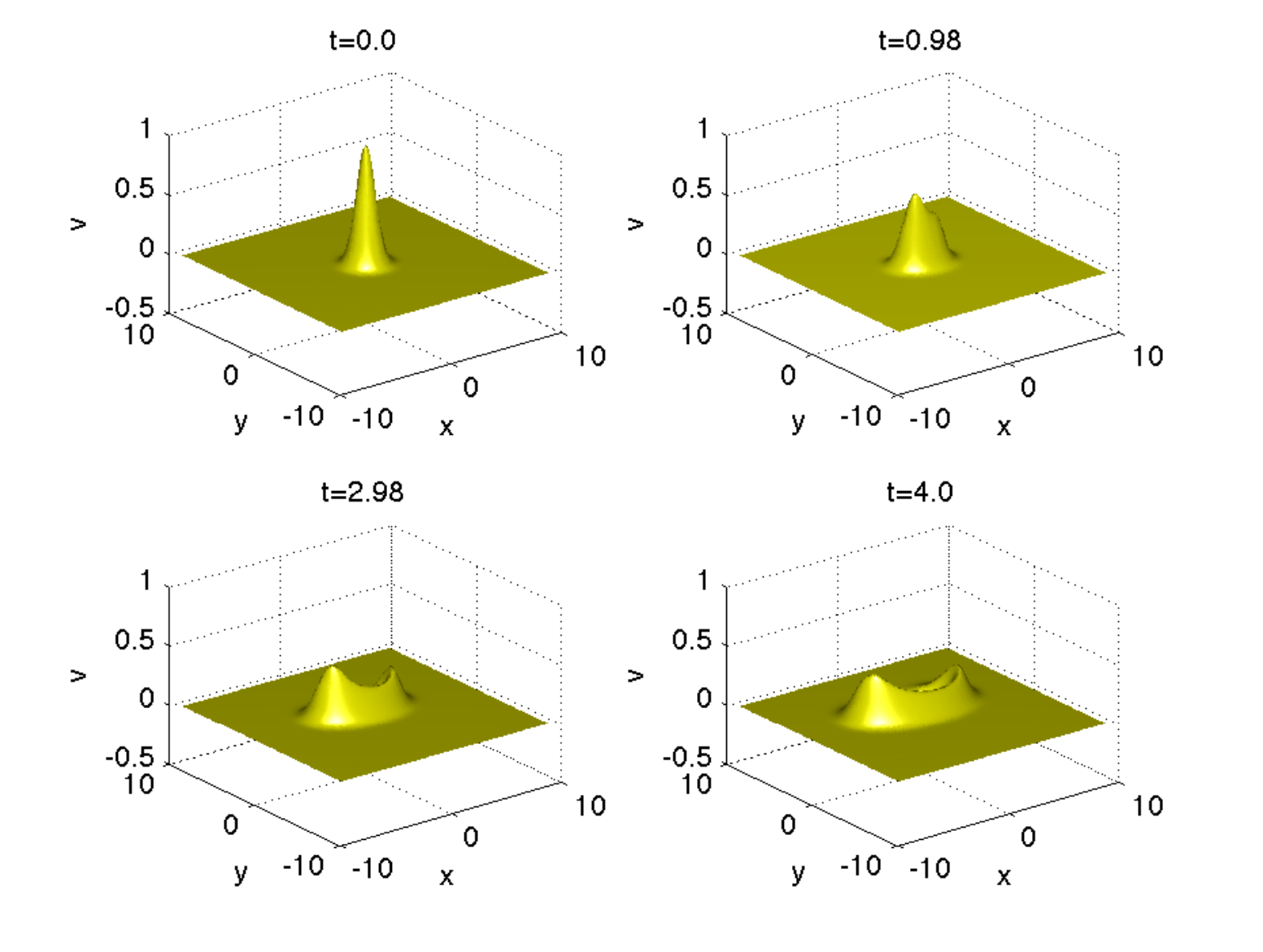} 
\caption{Solution $v$ to the system (\ref{max}) for the initial data 
$v(x,y,0)=\exp(-(x^{2}+y^{2}))$ and $w(x,y,0)=-1$ for several values 
of $t$.}
\label{vgauss_4td}
\end{center}
\vspace*{-5mm}
\end{figure}

A similar behavior can be observed for the 
solution $w$ in Fig.~\ref{wgauss_4td}. This solution starts with $w=-1$ 
and develops peaks above and below this initial surface.
It can be seen that both 
functions $v$ and $w$ approach an elliptic shape which could indicate the 
existence of an elliptic localized solution which acts as an attractor for a 
wide class of initial data. 
\begin{figure}
[!htbp]
\begin{center}
  \includegraphics[width=14cm]{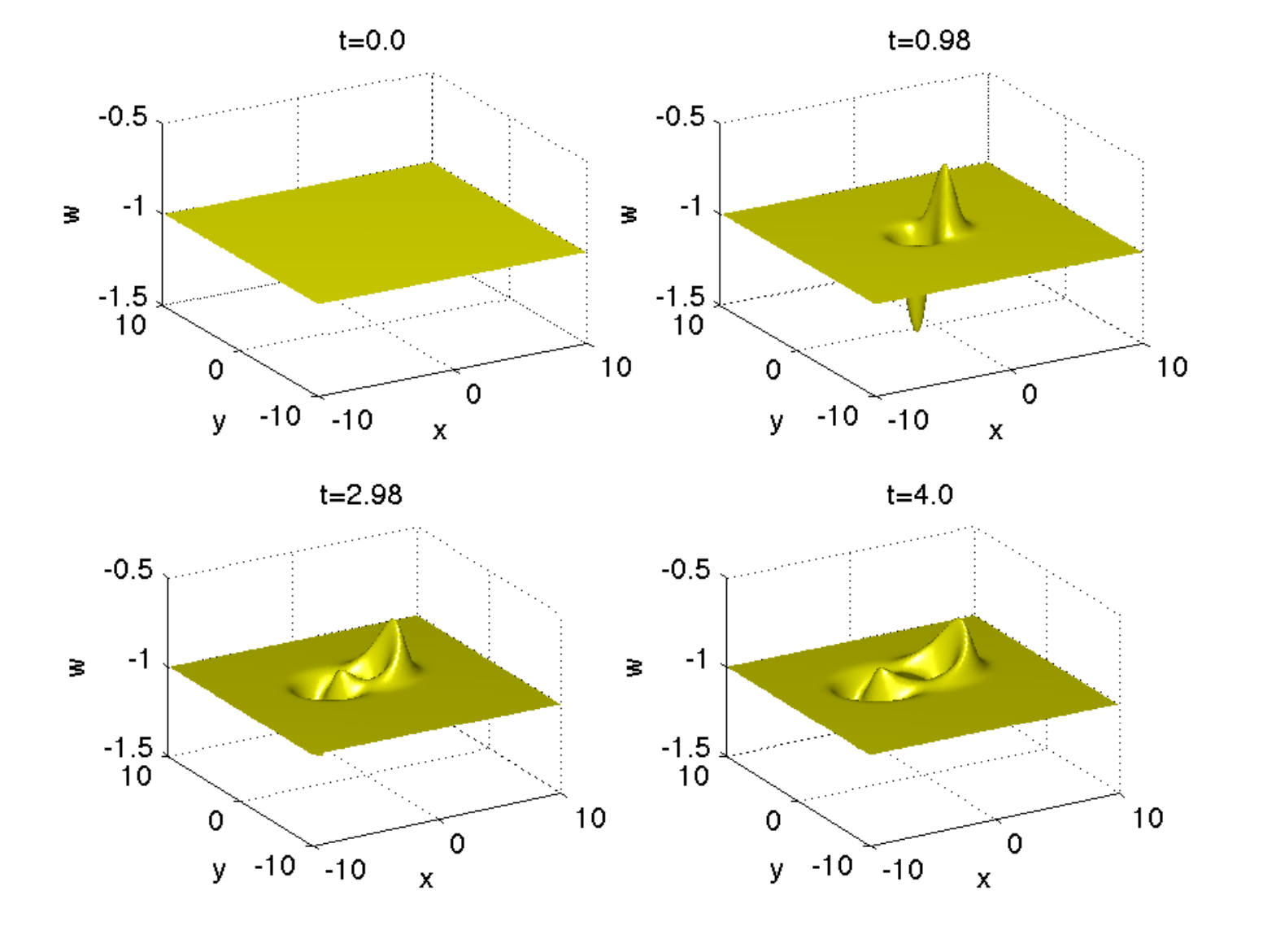} 
\caption{Solution $w$ to the system (\ref{max}) for the initial data 
$v(x,y,0)=\exp(-(x^{2}+y^{2}))$ and $w(x,y,0)=-1$ for several values 
of $t$.}
\label{wgauss_4td}
\end{center}
\vspace*{-5mm}
\end{figure}

The elliptic shape seems to be also present if one starts with 
several humps. In Fig.~\ref{vw2gausst4} we consider initial data of 
the form $v(x,y,0)=\exp(-((x-3)^{2}+y^{2}))+\exp(-((x+3)^{2}+y^{2}))$ 
and $w(x,y,0)=-1$. As can be seen both humps develop into an ellipse. 
The behavior for $w$ is as follows: since it starts as identically 
constant, two humps of opposite sign are forming. Both develop then 
an elliptic shape as seen in Fig.~~\ref{vw2gausst4}.
\begin{figure}[!htbp]
\begin{center}
  \includegraphics[width=14cm]{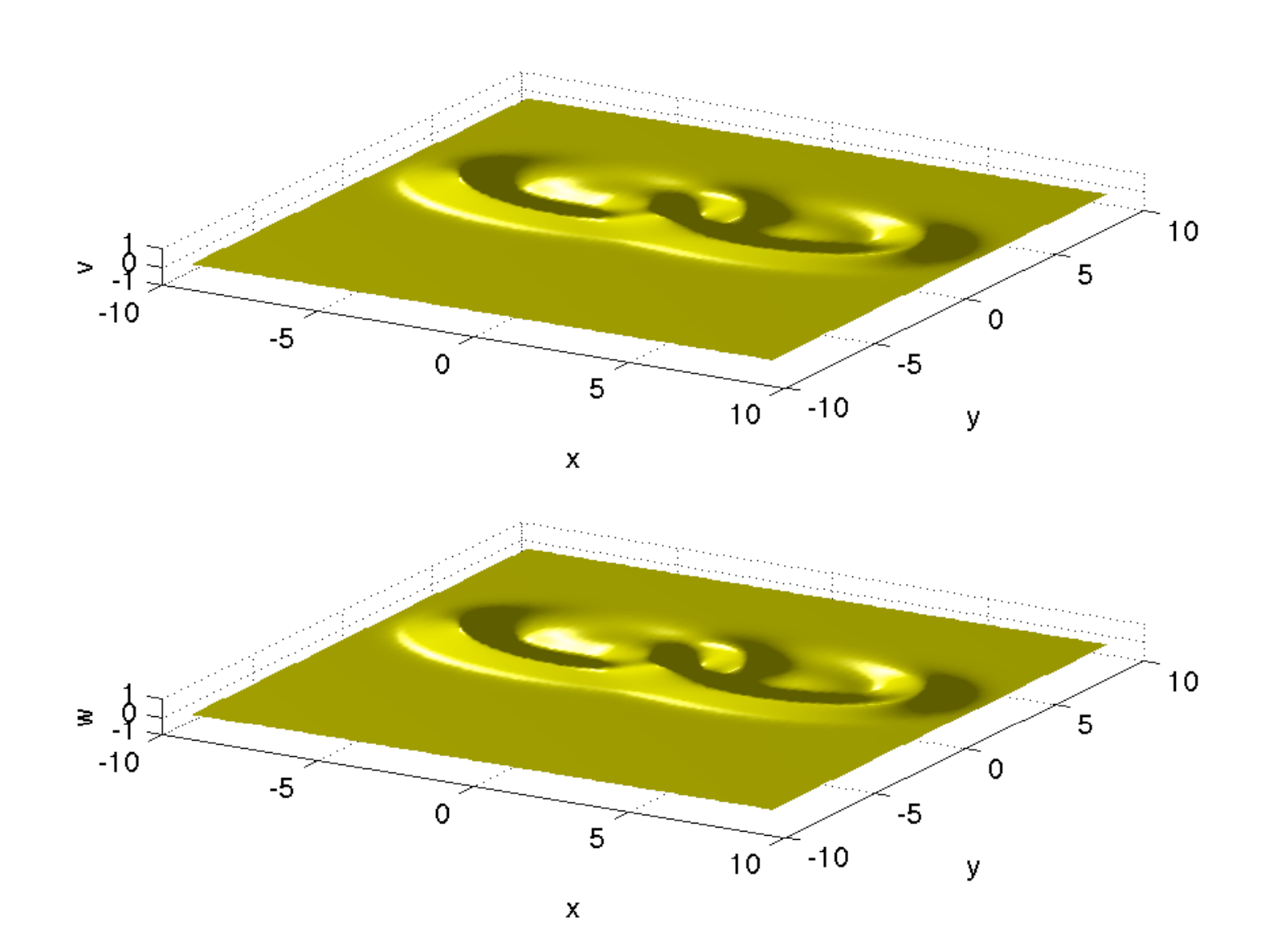} 
\caption{Solutions $v$ and $w$ to the system (\ref{max}) for the initial data 
$v(x,y,0)=\exp(-((x-3)^{2}+y^{2}))+\exp(-((x+3)^{2}+y^{2}))$ and $w(x,y,0)=-1$ for  $t=4$.}
\label{vw2gausst4}
\end{center}
\vspace*{-5mm}
\end{figure}

\section{Proof of Theorem 1}

In Sect. 4.1 we summarize the integrability conditions for two-component $(2+1)$-dimensional quasilinear systems. Their differential consequences are  used in Sect. 4.2 to establish Theorem 1. 

\subsection{Integrability conditions for $2$-component systems in $2+1$ dimensions}

Here we present the integrability conditions for $2$-component  quasilinear systems of the form (\ref{2+1}). Setting ${\bf u}=(v, w)^t$ and diagonalizing the matrix $A$, which is always possible in the $2\times 2$ case, one obtains the equations
\begin{equation}
\left(
\begin{array}{c}
v \\
\ \\
w
\end{array}
\right)_t+
\left(
\begin{array}{cc}
a & 0 \\
\ \\
0 & b
\end{array}
\right)
\left(
\begin{array}{c}
v \\
\ \\
w
\end{array}
\right)_x+
\left(
\begin{array}{cc}
p&q \\
\ \\
r&s
\end{array}
\right)
\left(
\begin{array}{c}
v \\
\ \\
w
\end{array}
\right)_y=0.
\label{sys}
\end{equation}
The  integrability conditions constitute a complicated set of second order differential constraints for the coefficients $a, b, p, q, r, s$ which were derived in \cite{Fer4} based on the method of hydrodynamic reductions.

\noindent {\bf Equations for $a$:}
\begin{eqnarray}
a_{vv}&=&\frac{\displaystyle{qa_vb_v+2qa_v^2+(s-p)a_va_w-ra_w^2}}{\displaystyle{(a-b)q}}+\frac{\displaystyle{a_vr_v}}{\displaystyle{r}}+\frac{\displaystyle{2a_vp_w-a_wp_v}}{\displaystyle{q}}, 
\nonumber \\
a_{vw}&=&a_v\frac{\displaystyle{a_w+b_w}}{\displaystyle{a-b}}+a_v(\frac{\displaystyle{q_w}}{q}+\frac{\displaystyle{r_w}}{r}), 
\label{a} \\
a_{ww}&=&\frac{\displaystyle{qa_vb_v+(s-p)a_vb_w+ra_w^2}}{\displaystyle{(a-b)r}}+\frac{\displaystyle{a_vs_w}}{r}+\frac{\displaystyle{a_wq_w}}{q}.
\nonumber
\end{eqnarray}

\noindent  {\bf Equations for $b$:}
\begin{eqnarray}
\nonumber \\
b_{vv}&=&\frac{\displaystyle{ra_wb_w+(p-s)a_vb_w+qb_v^2}}{\displaystyle{(b-a)q}}+\frac{\displaystyle{b_wp_v}}{q}+\frac{\displaystyle{b_vr_v}}{r}; 
\nonumber \\
b_{vw}&=&b_w\frac{\displaystyle{a_v+b_v}}{\displaystyle{b-a}}+b_w(\frac{\displaystyle{q_v}}{q}+\frac{\displaystyle{r_v}}{r}), 
\label{b} \\
b_{ww}&=&\frac{\displaystyle{ra_wb_w+2rb_w^2+(p-s)b_vb_w-qb_v^2}}{\displaystyle{(b-a)r}}+\frac{\displaystyle{b_wq_w}}{\displaystyle{q}}+\frac{\displaystyle{2b_ws_v-b_vs_w}}{\displaystyle{r}}.
\nonumber
\end{eqnarray}

\noindent  {\bf Equations for $p$:}
\begin{eqnarray}
p_{vv}&=&2\frac{r(a_vb_w-a_wb_v)+(s-p)a_vb_v}{(a-b)^2}+\frac{r_vp_v}{r}+\frac{p_vp_w}{q}+ 
\nonumber \\
&&\frac{\frac{r}{q} 
(2q_va_w-2a_vq_w+a_wp_w)-b_vp_v+2r_va_w-2a_v(s_v+p_v+r_w)+\frac{p-s}{q}(2p_va_w-a_vp_w)}{b-a}\nonumber 
\\
\ \nonumber \\
p_{vw}&=&2(s-p)\frac{a_vb_w}{(a-b)^2}-\frac{b_wp_v+(2s_w+p_w)a_v}{b-a}+{p_v}\left 
(\frac{q_w}{q}+\frac{r_w}{r}\right ), \label{p} \\
\  \nonumber \\
p_{ww}&=&2\frac{q(a_wb_v-a_vb_w)+(s-p)a_wb_w}{(a-b)^2}+\frac{(p-s)b_wp_v-qb_vp_v-2rs_wa_w-ra_wp_w}{(b-a)r}+ 
\nonumber \\
&& \frac{p_vs_w}{r}+\frac{q_wp_w}{q}. \nonumber
\end{eqnarray}

\noindent  {\bf Equations for $s$:}

\begin{eqnarray}
s_{vv}&=&2\frac{r(a_wb_v-a_vb_w)+(p-s)a_vb_v}{(a-b)^2}+\frac{(s-p)a_vs_w-ra_ws_w-2qp_vb_v-qb_vs_v}{(a-b)q}+ 
\nonumber \\
&&\frac{p_vs_w}{q}+\frac{r_vs_v}{r}, \nonumber \\
\  \nonumber \\
s_{vw}&=&2(p-s)\frac{a_vb_w}{(a-b)^2}-\frac{a_vs_w+(2p_v+s_v)b_w}{a-b}+{s_w}\left 
(\frac{q_v}{q}+\frac{r_v}{r}\right ), \label{s} \\
\  \nonumber \\
s_{ww}&=&2\frac{q(a_vb_w-a_wb_v)+(p-s)a_wb_w}{(a-b)^2}+\frac{q_ws_w}{q}+\frac{s_vs_w}{r} 
+ \nonumber \\
&&\frac{\frac{q}{r} 
(2r_wb_v-2b_wr_v+b_vs_v)-a_ws_w+2q_wb_v-2b_w(p_w+s_w+q_v)+\frac{s-p}{r}(2s_wb_v-b_ws_v)}{a-b}. 
\nonumber
\end{eqnarray}
\noindent {\bf Equations for  $q$ and $r$:}
\begin{eqnarray}
qr_{ww}+rq_{ww}&=&2(p-s)\frac{(p-s)a_wb_w+q(a_vb_w-a_wb_v)}{(a-b)^2} 
+q\frac{r_v}{r}\frac{qb_v+(s-p)b_w}{a-b}+   \nonumber \\
&&(s-p)\frac{2a_ws_w+2b_wp_w+b_wq_v}{a-b}+r\frac{(a_w-2b_w)q_w}{a-b}+ 
\nonumber \\
&&q\frac{a_wr_w+b_v(2p_w+2s_w+q_v)-2b_w(r_w+p_v+s_v)}{a-b}+ \nonumber \\
&&\frac{r}{q}q_w^2+\frac{q}{r}s_wr_v-q_wr_w+s_w(2p_w+q_v), \nonumber \\
\ \nonumber \\
q_{vw}&=&(s-p)\frac{qa_vb_v+(s-p)a_vb_w+ra_wb_w}{r(a-b)^2} 
+\frac{q_vq_w}{q}+\frac{p_vs_w}{r}+                     \nonumber \\
&& 
\frac{a_v(rq_w+qr_w)+(s-p)(a_vs_w+b_wp_v)+ra_ws_w+qp_vb_v}{r(a-b)}, 
\nonumber \\
\  \label{qr} \\
r_{vw}&=&(p-s)\frac{ra_wb_w+(p-s)a_vb_w+qa_vb_v}{q(a-b)^2} 
+\frac{r_vr_w}{r}+\frac{p_vs_w}{q}+                     \nonumber \\
&& 
\frac{b_w(rq_v+qr_v)+(p-s)(a_vs_w+b_wp_v)+ra_ws_w+qp_vb_v}{q(b-a)}, 
\nonumber \\
\ \nonumber \\
qr_{vv}+rq_{vv}&=&2(s-p)\frac{(s-p)a_vb_v+r(a_vb_w-a_wb_v)}{(a-b)^2} 
+r\frac{q_w}{q}\frac{ra_w+(p-s)a_v}{b-a}+   \nonumber \\
&&(p-s)\frac{2b_vp_v+2a_vs_v+a_vr_w}{b-a}+q\frac{(b_v-2a_v)r_v}{b-a}+ 
\nonumber \\
&&r\frac{b_vq_v+a_w(2s_v+2p_v+r_w)-2a_v(q_v+s_w+p_w)}{b-a}+ \nonumber \\
&&\frac{q}{r}r_v^2+\frac{r}{q}p_vq_w-r_vq_v+p_v(2s_v+r_w); \nonumber
\end{eqnarray}
These formulas are completely symmetric under the identification 
$v\leftrightarrow w, \ a \leftrightarrow b, \ p\leftrightarrow s, \ 
q\leftrightarrow r$.  Our next goal is to rewrite these equations in a somewhat different form which will be used in the proof of Theorem 1. First of all, Eqs.  $(\ref{a})_2$ and $(\ref{b})_2$ can be cast into the form
\begin{equation}
d\ln qr=\Omega, ~~~~~
\Omega=\left(\frac{b_{vw}}{b_w}+\frac{a_v+b_v}{a-b}  \right)\ 
dv+\left(\frac{a_{vw}}{a_v}+\frac{a_w+b_w}{b-a}  \right)\ dw.
\label{Omega}
\end{equation}
In what follows we assume that all quantities $a_v, a_w, b_v, b_w$ are nonzero, and parametrize  the first order derivatives of $q$ and $r$ as
\begin{equation}
\begin{array}{c}
\displaystyle \frac{q_{w}}{q}=\frac{a_{ww}}{a_w}+\frac{a_w}{b-a}-n, 
~~~~ \frac{r_{v}}{r}=\frac{b_{vv}}{b_v}+\frac{b_v}{a-b}-m, \\
\ \\
\displaystyle \frac{q_{v}}{q}=\frac{b_{vw}}{b_w}-\frac{b_{vv}}{b_v}+\frac{a_v}{a-b}+m, 
~~~~ \frac{r_{w}}{r}=\frac{a_{vw}}{a_v}-\frac{a_{ww}}{a_w}+\frac{b_w}{b-a}+n, 
\end{array}
 \label{qr1} 
\end{equation}
where $m$ and $n$ are two auxiliary functions. Solving the remaining equations (\ref{a}), (\ref{b})  for the first order derivatives of $p$ and $s$ one gets
\begin{eqnarray}
p_v&=&\frac{b_v}{b_w}qm+\frac{a_w}{a-b} r+\frac{a_v}{a-b}(p-s),   \nonumber \\
s_w&=&\frac{a_w}{a_v}rn+\frac{b_v}{b-a} q+\frac{b_w}{b-a}(s-p),   \nonumber \\
\  \label{ps1} \\
p_w&=&\frac{a_w}{b-a}(s-p)+\frac{a_w^2}{a_v(a-b)}r + 
\frac{1}{2}q\left( \left( \ln\frac{a_v}{b_v}\right)_v+\left(1+\frac{a_wb_v}{a_vb_w}\right)m+2\frac{a_v+b_v}{b-a}\right), \nonumber \\
s_v&=&\frac{b_v}{a-b}(p-s)+\frac{b_v^2}{b_w(b-a)}q + 
\frac{1}{2}r\left( \left( \ln\frac{b_w}{a_w}\right)_w+\left(1+\frac{a_wb_v}{a_vb_w}\right)n+2\frac{a_w+b_w}{a-b}\right). \nonumber \\
\nonumber
\end{eqnarray}
Substituting (\ref{ps1}) into the equations for $q_{vw}$ and $r_{vw}$ one gets
\begin{eqnarray}
{\ln q}_{vw}&=&\frac{a_{vw}}{a-b}-\frac{a_v(a_w+b_w)+a_wb_v}{(a-b)^2} + 
\frac{a_wb_v}{a_vb_w}mn+2\frac{a_w^2rn}{a_v(a-b)q}, \nonumber \\
\  \label{qr2} \\
{\ln r}_{vw}&=&\frac{b_{vw}}{b-a}-\frac{b_w(a_v+b_v)+a_wb_v}{(a-b)^2} + 
\frac{a_wb_v}{a_vb_w}mn+2\frac{b_v^2qm}{b_w(b-a)r}, \nonumber \\
\nonumber
\end{eqnarray}
respectively.  Finally, differentiating the equation $(\ref{ps1})_1$ with respect to $w$ and comparing the result with $(\ref{p})_2$, 
as well as differentiating  $(\ref{ps1})_2$ with respect to $v$ and comparing with $(\ref{s})_2$, we obtain
\begin{equation}
\begin{array}{c}
\displaystyle \left( \frac{b_v}{b_w}qm\right)_w-\frac{b_v}{b_w}qm\left( \frac{a_{vw}}{a_v}-\frac{a_w}{a-b}\right)+2n r \frac{a_w}{b-a}+3q\frac{a_vb_v}{(a-b)^2}=0,\\
\ \\
\displaystyle \left( \frac{a_w}{a_v}rn\right)_v-\frac{a_w}{a_v}rn\left( \frac{b_{vw}}{b_w}-\frac{b_v}{b-a}\right)+2mq \frac{b_v}{a-b}+3r\frac{a_wb_w}{(a-b)^2}=0.
\end{array}
\label{mn}
\end{equation}
Equations (\ref{Omega}), (\ref{qr2}) and (\ref{mn}) will be crucial for the proof of Theorem 1.

\subsection{ Proof of Theorem 1}

For definiteness, let us consider a two-component $(3+1)$-dimensional system of the form (\ref{1}). We assume that the generic matrix of the family  $\alpha^i A_i$ has a simple spectrum, and that the dispersion relation $\det(\mu I+\alpha^i A_i)=0$ defines an irreducible quadric. Diagonalizing the first matrix, we obtain the equations
$$
{\bf u}_t+ A({\bf u}) {\bf u}_{x}+B({\bf u}) {\bf u}_{y}+\tilde B({\bf u}) {\bf u}_{z}=0
$$
where ${\bf u}=(v, w)^t$, and the matrices $A, B, \tilde B$ have the form
$$
A=\left(
\begin{array}{cc}
a & 0 \\
\ \\
0 & b
\end{array}
\right), 
~~~ B=\left(
\begin{array}{cc}
p&q \\
\ \\
r&s
\end{array}
\right), ~~~ \tilde B=
\left(
\begin{array}{cc}
\tilde p&\tilde q \\
\ \\
\tilde r&\tilde s
\end{array}
\right).
$$
Due to the irreducibility of the dispersion relation, one cannot have $q=\tilde q=0$ or $r=\tilde r=0$. Thus, by taking appropriate linear combinations of $B$ and $\tilde B$, one can assume  $q, r, \tilde q, \tilde r$ to be nonzero. Since all traveling wave reductions of our system must be integrable, the matrices $B$,  $\tilde B$, and $B+\tilde B$ must satisfy the $(2+1)$-dimensional integrability conditions from Sect. 4.1 (with the same matrix $A$). 
There are three cases  to consider:

\noindent {\bf Case 1. } A generic   matrix of the family $\alpha A +\beta B + \tilde \beta \tilde B$ has constant eigenvalues. In particular,  all coefficients of the dispersion relation must be constant.  This implies that $a, b, p, s, \tilde p, \tilde s, qr, \tilde q\tilde r, q\tilde r+ r \tilde q$ must be constant, and the integrability conditions of Sect. 4.1, applied to pairs $(A, B)$,  $(A, \tilde B)$ and $(A, B+\tilde B)$, show that any such system must be  linearizable by a transformation of the form $v\to f(v), \ w \to g(w)$. 

\noindent {\bf Case 2. } A generic   matrix of the family $\alpha A +\beta B + \tilde \beta \tilde B$ is linearly degenerate. Then, due to the irreducibility of the dispersion relation, one cannot have a situation when a generic matrix of this family has one constant and one non-constant eigenvalue. Thus, we can assume that the matrix $A$ is linearly degenerate, and has non-constant eigenvalues.  In this case $A$ can be brought to the form (\ref{lindeg}), and both matrices $B$ and $\tilde B$ must be of the form (\ref{lindeg1}): see Example 1 of Sect. 2.  Any such system is therefore a traveling wave reduction of the system (\ref{max}).

\noindent {\bf Case 3. } A generic   matrix of the family $\alpha A +\beta B + \tilde \beta \tilde B$ is genuinely nonlinear (again, due to the irreducibility of the dispersion relation, one cannot have a `mixed' situation when a generic matrix of the family has one linearly degenerate, and one genuinely nonlinear eigenvalue). In particular, $a_v\ne 0$ and $b_w\ne 0$. In this case we will prove that matrices $A, B$ and $\tilde B$ must be linearly dependent, so that such system cannot be essentially $(3+1)$-dimensional.   Applying the condition (\ref{Omega}) to the matrices $B$, $\tilde B$ and $B+\tilde B$ one obtains
$$
d\ln qr=d\ln \tilde q\tilde r=d\ln (q+\tilde q)(r+\tilde r),
$$
so that $\tilde q=\lambda q, \ \tilde r = \mu r$ where $\lambda$ and $\mu$ are constants. In what follows we  assume that $a_w\ne 0$ and $b_v\ne 0$ (if both of these derivatives vanish,  one can take $a=v, b=w$, however, with such $A$ the integrability conditions from Sect. 4.1 are not compatible; similarly, due to the irreducibility of the dispersion relation one cannot have a situation when only one of these derivatives vanishes). If  $\lambda=\mu$, the equations (\ref{ps1}) imply
$\tilde p=\lambda p + \tau a +\eta, \  \tilde s=\lambda s + \tau b +\eta$,  so that
$\tilde B=\lambda B+ \tau A +\eta I$, and our system  becomes essentially $(2+1)$-dimensional. Thus, we assume $\lambda\ne \mu$. This means that, whenever one obtains a corollary of the integrability conditions which does not depend explicitly on $p$ and $s$,  and involves  $a, b, q, r$ only,  it should still hold after the rescaling $q\to \lambda q, \ r\to \mu r$. In particular, rescaling in this way the relations (\ref{qr2}) one obtains $m=n=0$. This, however, is inconsistent with relations (\ref{mn}): the last term  does not contain $m,n$, and cannot vanish. This finishes the proof.

\section{Concluding remarks}

Our results suggest that there must be a restriction on the dimensions in which integrable $n$-component systems can live. Namely, for  any number of components $n$, there must be an integer $d(n)$ such that there exist no $n$-component integrable systems of hydrodynamic type in the dimensions higher than $d(n)+1$. We  proved that $d(2)=5$, moreover, there exists a unique two-component $(5+1)$-dimensional integrable system.  It would be of interest to find a general formula for $d(n)$, and classify the systems which attain an upper bound of this inequality.

\section*{Acknowledgements}
We thank Boris Dubrovin and Maxim Pavlov for useful discussions. EVF thanks Institut de Math\'ematiques de Bourgogne for a financial support making this collaboration possible.
This work has been supported by the project FroM-PDE funded by the European
Research Council through the Advanced Investigator Grant Scheme. CK
thanks for financial support by the Conseil R\'egional de Bourgogne
via a FABER grant and the ANR via the program ANR-09-BLAN-0117-01.

\end{document}